\documentclass[12pt]{article}
\usepackage{graphicx}
\usepackage{authblk}
\usepackage{orcidlink,amsmath,rotating,amssymb,amsthm,algorithm,epstopdf,multicol,multirow,gensymb,subcaption,fontenc,algorithmic,tablefootnote,longtable,array}
\usepackage[section]{placeins}
\usepackage{enumerate}
\usepackage{natbib}
\usepackage{url} % not crucial - just used below for the URL 
\usepackage{hyperref}
\usepackage{xcolor,soul,chngcntr}
\hypersetup{colorlinks=true,linkcolor=,citecolor=blue}

\newcommand{\blind}{1}

\begin{document}

\def\spacingset#1{\renewcommand{\baselinestretch}%
{#1}\small\normalsize} \spacingset{1}

%%%%%%%%%%%%%%%%%%%%%%%%%%%%%%%%%%%%%%%%%%%%%%%%%%%%%%%%%%%%%%%%%%%%%%%%%%%%%%

\if1\blind
{
  \title{\bf Nonparametric contaminated Gaussian mixture of regressions}
  \author{Sphiwe B. Skhosana\textsuperscript{1*}\orcidlink{0000-0002-4740-7540} and Weixin Yao\textsuperscript{2}\vspace{.5cm}\\
\textsuperscript{1}Department of Statistics, University of Pretoria, Pretoria, 0028, South Africa\vspace{.5cm}\\
\textsuperscript{2}Department of Statistics, University of California, Riverside, CA, USA\vspace{.5cm}\\
    \textsuperscript{*}Corresponding author: \url{spiwe.skhosana@up.ac.za}}
  \maketitle
} \fi

\if0\blind
{
  \bigskip
  \bigskip
  \bigskip
  \begin{center}
    {\LARGE\bf Nonparametric contaminated Gaussian mixture of regressions}
\end{center}
  \medskip
} \fi

\bigskip
\begin{abstract}
Semi- and non-parametric mixture of regressions are a very useful flexible class of mixture of regressions in which some or all of the parameters are non-parametric functions of the covariates. These models are, however, based on the Gaussian assumption of the component error distributions. Thus, their estimation is sensitive to outliers and heavy-tailed error distributions. In this paper, we propose semi- and non-parametric contaminated Gaussian mixture of regressions to robustly estimate the parametric and/or non-parametric terms of the models in the presence of mild outliers. The virtue of using a contaminated Gaussian error distribution is that we can simultaneously perform model-based clustering of observations and model-based outlier detection. We propose two algorithms, an expectation-maximization (EM)-type algorithm and an expectation-conditional-maximization (ECM)-type algorithm, to perform maximum likelihood and local-likelihood kernel estimation of the parametric and non-parametric of the proposed models, respectively. The robustness of the proposed models is examined using an extensive simulation study. The practical utility of the proposed models is demonstrated using real data.
\end{abstract}

\noindent%
{\it Keywords: Expectation-Maximization (EM) algorithm, Non-parametric regression, Mixture of regressions, contaminated Gaussian}
\vfill

\newpage
\spacingset{1.4} % DON'T change the spacing!

\section{Introduction}\label{sec1}
Finite mixtures of regression models are very useful for studying the relationship between a response variable $y$ and a set of covariates $\mathbf{x}=(x_1,x_2,\dots,x_p)^\top$ when the underlying population is made up of a, typically unknown, number of, say $K$, unobserved sub-populations also known as components.\\
Let $Z$ be a component indicator variable with a discrete distribution $P(Z=k)=\pi_k$, for $k=1,2,\dots,K$, where each $\pi_k>0$ and $\sum^{K}_{k=1}\pi_k=1$. Given that $Z=k$ and $\mathbf{X}=\mathbf{x}$, the relationship between $y$ and $\mathbf{x}$ is given by
\begin{equation}
    y=m_k(\mathbf{x})+\epsilon_k,\text{ for } k=1,2,\dots,K
\end{equation}
where $m_k(\mathbf{x})$ is the regression function, $\epsilon_k$ is the error term, typically assumed to: (1) be independent of $\mathbf{x}$ and (2) follow a Gaussian distribution with zero mean and variance $\sigma^2_k$. That is $\epsilon_k\sim \mathcal{N}(0,\sigma^2_k)$, where $\mathcal{N}(\mu,\sigma^2)$ denotes the Gaussian distribution with mean $\mu$ and variance $\sigma^2$. Since $Z$ is typically unobserved, given $\mathbf{X}=\mathbf{x}$, $y$ is said to follow a mixture of Gaussian distributions
\begin{eqnarray}
    f(y|\mathbf{X}=\mathbf{x})&=&\pi_1\mathcal{N}\{y|m_1(\mathbf{x}),\sigma^2_1\}+\dots+\pi_K\mathcal{N}\{y|m_K(\mathbf{x}),\sigma^2_K\}\\
    &=&\sum_{k}^K\pi_k\mathcal{N}\{y|m_k(\mathbf{x},\sigma^2_k\}\label{model1}
\end{eqnarray}
where the $\pi_k$'s are the mixing probabilities specifying the relative size of each component. Model \eqref{model1} is a finite Gaussian mixture of regressions (GMRs) model.\\
If $m_k(\mathbf{x})=\mathbf{x}^\top\boldsymbol{\beta}_k$, where $\mathbf{x}=(1,x_1,x_2,\dots,x_p)^\top$ and $\boldsymbol{\beta}_k=(\beta_{k0},\beta_{k1},\dots,\beta_{kp})$ is the regression parameter vector, model \eqref{model1} is a finite Gaussian mixture of linear regressions (GMLRs) first introduced and studied by \cite{quandt1972} and subsequently by \cite{quandt1978} as switching regressions models in economics. Subsequent to their introduction in economics, GMLRs received widespread use in marketing \cite{desarbo1988}, machine learning \cite{jacobs1991}, environmental economics \cite{huang2013} and medicine \cite{schlattmann2009}, among many other fields. See Chapter 8 of \cite{frühwirth2006} for a comprehensive overview of the theory and application of GMLRs and Chapter 3 of \cite{yao2024} for a recent review of GMLRs and extensions.\\
The parametric assumption imposed by GMLRs, that is, constant variances and mixing probabilities as well as the linear regression functions are quite restrictive and usually fail to hold in practice. For this reason, \cite{huang2013} proposed to relax these assumptions by introducing non-parametric GMRs (NPGMRs)
\begin{eqnarray}
f(y|\mathbf{X}=\mathbf{x})=\sum_{k=1}^K\pi_k(\mathbf{x})\mathcal{N}\{y|m_k(\mathbf{x}),\sigma^2_k(\mathbf{x})\}. \label{model2}
\end{eqnarray}
Model \eqref{model2} allows the mixing probabilities, regression functions and variances to be functions of the covariates $\mathbf{x}$. Due to the curse of dimensionality, the authors considered only the case when $\mathbf{x}$ is univariate, hence $X=x$ in model \eqref{model2}. Although highly flexible, model \eqref{model2} suffers from a loss of efficiency in estimation. For this reason, \cite{xiang2018} proposed the semi-parametric GMRs (SPGMRs)
\begin{eqnarray}
f(y|X=x)=\sum_{k=1}^K\pi_k\mathcal{N}\{y|m_k(x),\sigma^2_k\}. \label{model3}
\end{eqnarray}
Model \eqref{model3} assumes that $\pi_k$ and $\sigma_k^2$ are constant, as in the GMLRs, whereas $m_k(\cdot)$ is a non-parametric function of $x$ as in model \eqref{model2}.\\
Note that the models discussed do far assume that each component has a Gaussian distribution. Unfortunately, in practice, data may be contaminated by outliers which may have a significant negative effect on the model estimation \cite{mazza2020}. Mazza and Punzo \cite{mazza2020} distinguish between mild and gross outliers. The authors define mild outliers as a points that do not significantly deviate from the regression relationship within a given component, however, these points produce component distributions that are too heavy-tailed to be adequately modelled by a Gaussian distribution. On the other hand, gross outliers are points that are significantly away from any of the regression components. Our focus in this paper is on mild outliers. For more details about the distinction between these two classes of outliers, see \cite{mazza2020} and the references therein.\\
There are a number of approaches that have been proposed to obtain robust estimates of a mixture of regressions model in the presence of outliers. One class of methods to robustly estimate the mixture of regressions model is to modify the model estimation procedure by trimming (see, \cite{neykov2007}, \cite{garcia2017}, \cite{dougru2018}), assigning weights to different data points (\cite{markatou2000}, \cite{shen2004}) or modifying the fitting algorithm (\cite{bai2012}). See the review \cite{yu2020} for more details about these approaches. Another simple class of methods to handle outliers in mixture of regression modelling is to replace the Gaussian component error distribution with a heavy-tailed distribution. This will be our focus in this paper. We first note the proposals that have been made along these lines. Yao \emph{et al.} \cite{yao2014} proposed a mixture of linear regressions where each component error distribution follows a $t-$distribution. Song \emph{et al.} \cite{song2014} made a similar proposal, however they assumed that each component error distribution follows a Laplace distribution. Recently, \cite{mazza2020} proposed mixtures of linear regressions with a contaminated Gaussian distribution (hereafter, CGMLRs), where each component error distribution follows a contaminated Gaussian distribution. This is a special case of the more general model proposed by Zeller \emph{et al.} \cite{zeller2016} to model both skewness and outliers using scale mixture of skew normal error component distributions. In contrast to the models proposed by \cite{yao2014} and \cite{song2014}, the CGMLRs can simultaneously cluster an observation into a group or component and thereafter decide whether or not the observation is an outlier using a probabilistic approach.\\
For this reason, in this paper, we propose non-parametric contaminated Gaussian mixture of regressions (NPCGMRs) and semi-parametric contaminated Gaussian mixture of regressions (SPCGMRs). These models arises as a result of combining the flexibility of NPGMRs \eqref{model2} and SPGMRs \eqref{model3} with the dual ability of CGMLRs for simultaneous probabilistic clustering and outlier detection.\\
The rest of the paper is organised as follows. In Section \ref{NPCGMRs}, we introduce the proposed NPCGMRs model, we discuss the conditions under which the model is identifiable and propose a local-likelihood estimation procedure via a modified Expectation-Maximization (EM) algorithm to estimate the model. In Section \ref{SPCGMRs}, we introduce the proposed SPCGMRs model, we discuss its identifiability followed by the proposed estimation procedure for the model. In Section \ref{model-based} and \ref{model-select}, we discuss model-based cluster and automatic outlier detection using the proposed models. In Section \ref{model-select}, we discuss how to select the number of mixture components for the proposed models. In Section \ref{simulations}, we conduct extensive Monte Carlo simulation to demonstrate the performance and practical use of the proposed models and estimation procedure. The proposed models are applied to real data in Section \ref{app}. Section \ref{conc} concludes the paper and gives directions for future studies. 
\section{Non-parametric contaminated Gaussian mixtures of regressions (NPCGMRs)}\label{NPCGMRs}
In this section, we define the NPCGMRs model, state the conditions under which the model is identifiable followed by an estimation procedure for the model. 
\subsection{Model definition and Identifiability}
Consider a random sample $\{(x_i,y_i):i=1,2,\dots,n\}$ from the population $(X,Y)$, where $Y$ is the response variable whose variation is explained by the covariate $X$. To aid the reader's comprehension of the proposed methods and partly to avoid complications with the "curse of dimensionality", we consider the case of a single covariate. However, the proposed methods can be extended to the case of more than one covariate. We give directions towards this end in our future studies.\\
Let $Z$ be a component indicator with a conditional discrete distribution $P(Z=k|X=x)=\pi_k(x)$, for $k=1,2,\dots,K$, where $\pi_k(x)>0$ and $\sum^K_k\pi_k(x)=1$ for all $x$. Conditional on $Z=k$ and $X=x$, $y$ is assumed to follow a contaminated Gaussian distribution with density function
\begin{equation}
    f_{CG}(y|X=x;\theta_k)=\alpha_k\mathcal{N}\{y|m_k(x),\sigma_k^{2}(x)\}+(1-\alpha_k)\mathcal{N}\{y|m_k(x),\eta_k\sigma_k^2(x)\}
\end{equation}
where $\alpha_k\in(0,1)$ is the proportion of good points, $\eta_k>1$ is the degree of contamination. Because of the assumption $\eta_k>1$, $\eta_k$ can be interpreted as the increase in the variability due to the presence of outliers. Lastly, $\boldsymbol{\theta}_k=(\alpha_k,\mathbf{m}_k,\boldsymbol{\sigma}_k^2,\eta_k)$, where $\mathbf{m}_k=\{m_k(x)|x\in \mathcal{X}\}$ and $\boldsymbol{\sigma}^2_k=\{\sigma^2_k(x)|x\in \mathcal{X}\}$, and $\mathcal{X}$ denotes the domain of $x$.\\
Given only $X=x$, $y$ is said to follow a mixture of contaminated Gaussian distributions with density function
\begin{align}
p(y|X=x;\boldsymbol{\Theta})&=\sum_{k}^K\pi_k(x)f_{CG}(y|X=x;\theta_k)\nonumber\\
    &=\sum_{k=1}^K\pi_k(x)\big[\alpha_k\mathcal{N}\{y|m_k(x),\sigma_k^{2}(x)\}+(1-\alpha_k)\mathcal{N}\{y|m_k(x),\eta_k\sigma_k^2(x)\}\big]\label{model4}
\end{align}
We assume that $\pi_k(\cdot)$, $m_k(\cdot)$ and $\sigma^2_k(\cdot)$ are unknown but smooth functions of the covariate $x$, hence non-parametric. $\boldsymbol{\Theta}=(\Theta_1,\Theta_2,\dots,\Theta_K)$, where $\Theta_k=(\boldsymbol{\pi}_k,\boldsymbol{\theta}_k)$, with $\boldsymbol{\pi}_k=\{\pi_k(x)|x\in \mathcal{X}\}$, is the vector of all the parametric and non-parametric terms of the model. For a given value of $K$, model \eqref{model4} is a finite non-parametric contaminated Gaussian mixture of regressions (NPCGMRs).\\
The identifiability of any mixture model is a necessary requirement for model estimation and inference. \cite{huang2013} and \cite{wang2014} showed that the NPGMRs is identifiable if (i) $\pi_k(x)>0$ is a continuous function, and $m_k(x)$ and $\sigma^2_k(x)$ are differentiable functions, for $k=1,2,\dots,K$; (ii) the domain $\mathcal{X}$ of $x$ is an interval in $\mathbb{R}$; (iii) for any two sets of functions $(m_i(x),\sigma^2_i(x))$ and $(m_j(x),\sigma^2_j(x))$, for $i\neq j$,
\begin{equation}
    \sum^{1}_{i=1}||m^{(1)}_i(x)-m^{(1)}_j(x)||^2+||\sigma^{2(1)}_i(x)-\sigma^{2(1)}_j(x)||^2\neq 0
\end{equation}
and (iv) the parametric mixture model $\sum_{k=1}^K\pi_k\mathcal{N}\{y;\mu_k,\sigma^2_k\}$ is identifiable.\\
We assume that the first three conditions (i) - (iii) hold for the proposed NPCGMRs model. This is a reasonable assumption. Condition (iv) follows from \cite{punzo2016}. Then, the NPCGMRs model is identifiable.
\subsection{Model estimation}\label{backfit1}
In this section, we propose an estimation procedure for estimating model \eqref{model4} and an EM-type algorithm to implement the estimation procedure. 
\subsubsection{Estimation procedure}
Consider a random sample $\{(x_i,y_i):i=1,2,\dots,n\}$ from the model \eqref{model4}. The corresponding log-likelihood function is 
\begin{equation}
    \ell\{\boldsymbol{\Theta}\}=\sum_{i=1}^n\log\bigg[\sum_{k=1}^K\pi_k(x_i)\big(\alpha_k\mathcal{N}\{y|m_k(x_i),\sigma_k^{2}(x_i)\}+(1-\alpha_k)\mathcal{N}\{y|m_k(x_i),\eta_k\sigma_k^2(x_i)\}\big)\bigg]\label{loglik1}
\end{equation}
Note that $\boldsymbol{\Theta}$ is made up of parametric terms $\{(\alpha_k,\eta_k):k=1,2,\dots,K\}$ and non-parametric terms $\{(\pi_k(x_i),m_k(x_i),\sigma^2_k(x_i)):i=1,2,\dots,n;k=1,2,\dots,K\}$. Thus, the log-likelihood function \eqref{loglik1} cannot be directly maximized. As in \cite{xiang2018} and \cite{xiang2020}, we propose a backfitting algorithm which iterates between estimating the parametric terms and the non-parametric terms.\\
Given the estimates $\{(\hat{\alpha}_k,\hat{\eta}_k):k=1,2,\dots,K\}$ of the parametric terms, we estimate the non-parametric terms using local-likelihood estimation \cite{tibshirani1987}. Let $u\in \mathcal{U}$, where $\mathcal{U}$ is the set of all local points in the domain $\mathcal{X}$ of the covariate $x$. Thus $\mathcal{U}\subset \mathcal{X}$. We make use of the local-constant estimator to obtain $\{(\hat{\pi}_k(u),\hat{m}_k(u),\hat{\sigma}^2_k(u)):u\in \mathcal{U};k=1,2,\dots,K\}$ by maximizing the locally weighted log-likelihood function, over all $u\in \mathcal{U}$,
\begin{eqnarray}
    \ell^1\{\boldsymbol{\Theta}(u)\}=\sum_{i=1}^n\log\bigg[\sum_{k=1}^K\pi_k(u)\big(\hat{\alpha}_k\mathcal{N}\{y|m_k(u),\sigma_k^{2}(u)\}+\nonumber\\
    (1-\hat{\alpha}_k)\mathcal{N}\{y|m_k(x_i),\hat{\eta}_k\sigma_k^2(u)\}\big)\bigg]W_h(x_i-u)\label{loglik2}
\end{eqnarray}
where $\boldsymbol{\Theta}(u)=\{(\pi_k(u),m_k(u),\sigma^2_k(u)):k=1,2,\dots,K\}$ and $W_h(t)=W(t/h)/h$ is the rescaled kernel function with bandwidth $h$. Let $\{\hat{\boldsymbol{\Theta}}(u):u\in\mathcal{U}\}$ be the estimate of $\{\boldsymbol{\Theta}(u):u\in \mathcal{U}\}$. To obtain $\{\hat{\boldsymbol{\Theta}}(x_i): i=1,2,\dots,n\}$, the estimated non-parametric functions, for $x_i\notin \mathcal{U}$, we make use of interpolation.\\
Given the estimates $\{\hat{\boldsymbol{\Theta}}(x_i):i=1,2,\dots,n\}$, we update the parameter estimates $\{(\hat{\alpha}_k,\hat{\eta}_k):k=1,2,\dots,K\}$ by maximizing 
\begin{eqnarray}
    \ell^2\{\boldsymbol{\alpha},\boldsymbol{\eta}\}=\sum_{i=1}^n\log\bigg[\sum_{k=1}^K\hat{\pi}_k(x_i)\big(\alpha_k\mathcal{N}\{y|\hat{m}_k(x_i),\hat{\sigma}_k^{2}(x_i)\}+\nonumber\\
    (1-\alpha_k)\mathcal{N}\{y|\hat{m}_k(x_i),\eta_k\hat{\sigma}_k^2(x_i)\}\big)\bigg]\label{loglik3}
\end{eqnarray}
where $\boldsymbol{\alpha}=\{\alpha_k:k=1,2,\dots,K\}$ and $\boldsymbol{\eta}=\{\eta_k:k=1,2,\dots,K\}$. Let $\tilde{\boldsymbol{\alpha}}$ and $\tilde{\boldsymbol{\eta}}$ be the parameter estimates obtained from maximizing $\ell^2\{\cdot\}$.
\subsubsection{Fitting algorithm}
We propose two modified EM-type fitting algorithms to implement the above backfitting estimation procedure. 
\paragraph{Algorithm 1}
In order to obtain the initial parameter estimates $\{(\hat{\alpha}_k,\hat{\eta}_k):k=1,2,\dots,K\}$, we make use of the contaminated Gaussian mixture of linear regressions (CGMLRs) model proposed by \cite{mazza2020}. Given $\{(\hat{\alpha}_k,\hat{\eta}_k):k=1,2,\dots,K\}$, we maximize $\ell^1$ using a modified version of the classical expectation-maximization (EM) algorithm \cite{DLR1977} first proposed by \cite{huang2012}. Towards that end we introduce two latent variables. The first latent variable is $\mathbf{z}_i=(z_{i1},z_{i2},\dots,z_{iK})^\top$, for $i=1,2,\dots,n$, where $z_{ik}=1$ if $(x_i,y_i)$ belongs in the $k^{th}$ component and $0$ otherwise. The second latent variable is $\mathbf{v}_i=(v_{i1},v_{i2},\dots,v_{iK})^\top$, for $i=1,2,\dots,n$, where $v_{ik}=1$ if $(x_i,y_i)$ is not an outlier in the $k^{th}$ component and $0$ otherwise. The complete-data are $\{(x_i,y_i,\mathbf{z}_i,\mathbf{v}_i):i=1,2,\dots,n\}$ and the complete-data local log-likelihood is
\begin{eqnarray}
    \ell^1_c\{\boldsymbol{\Theta}(u)\}=\sum_{i=1}^n\sum_{k=1}^Kz_{ik}\log\bigg[\pi_k(u)\big(\hat{\alpha}_k\mathcal{N}\{y|m_k(u),\sigma_k^{2}(u)\})^{v_{ik}}\nonumber\\
    \times((1-\hat{\alpha}_k)\mathcal{N}\{y|m_k(u),\hat{\eta}_k\sigma_k^2(u)\}\big)^{1-v_{ik}}\bigg]W_h(x_i-u)\label{loglik4}
\end{eqnarray}
which can be written as
\begin{eqnarray}
     \ell^1_c\{\boldsymbol{\Theta}(u)\}= \ell^1_{1c}\{\boldsymbol{\pi}(u)\}+ \ell^1_{2c}\{\mathbf{m}(u),\boldsymbol{\sigma}^2(u)\}+\text{const}
\end{eqnarray}
where $\boldsymbol{\pi}(u)=\{\pi_k(u):k=1,2,\dots,K\}$, $\mathbf{m}(u)=\{m_k(u):k=1,2,\dots,K\}$, $\boldsymbol{\sigma}^2(u)=\{\sigma^2_k(u):k=1,2,\dots,K\}$,
\begin{eqnarray}
\ell^1_{1c}\{\boldsymbol{\pi}(u)\}&=&\sum_{i=1}^n\sum_{k=1}^Kz_{ik}\log\pi_k(u)W_h(x_i-u),
\end{eqnarray}
\begin{eqnarray}
\ell^1_{2c}\{\mathbf{m}(u),\boldsymbol{\sigma}^2(u)\}=\sum_{i=1}^n\sum_{k=1}^Kz_{ik}\big[{v_{ik}}\log\mathcal{N}\{y|m_k(u),\sigma_k^{2}(u)\})+\nonumber\\({1-v_{ik}})\log\mathcal{N}\{y|m_k(u),\hat{\eta}_k\sigma_k^2(u)\}\big)\big]W_h(x_i-u)
\end{eqnarray}
and ``const" represents all the terms that are independent of $\boldsymbol{\Theta}(u)$. The EM algorithm iterates between an expectation (E-) step, in which the conditional expectation of $\ell_c^1\{\boldsymbol{\Theta}(u)\}$, denoted by $Q\{\boldsymbol{\Theta}(u)|\boldsymbol{\Theta}^{(r-1)}(u)\}$, is calculated and a maximization (M-) step to update $\boldsymbol{\Theta}^{(r-1)}(u)$, for $u\in \mathcal{U}$, by maximizing $Q\{\boldsymbol{\Theta}(u)|\boldsymbol{\Theta}^{(r-1)}(u)\}$, where $r$ denotes the iteration number. At the $r^{th}$ iteration, the E-step reduces to the calculation of the conditional expectations $\mathbb{E}\big[z_{ik}|x_i,y_i,\boldsymbol{\Theta}^{(r-1)}(u)\big]=\gamma^{(r)}_{ik}(u)$ and  $\mathbb{E}\big[v_{ik}|x_i,y_i,\mathbf{z}_i,\boldsymbol{\Theta}^{(r-1)}(u)\big]=\lambda_{ik}^{(r)}(u)$, for $i=1,2,\dots,n$. As in \cite{mazza2020}, $\gamma^{(r)}_{ik}(u)$ is the posterior probability that the $i^{th}$ data point belongs to the $k^{th}$ and $\lambda^{(r)}_{ik}(u)$ is the posterior probability that the $i^{th}$ data point is not an outlier in the $k^{th}$ component. To obtain $Q\{\boldsymbol{\Theta}(u)|\boldsymbol{\Theta}^{(r-1)}(u)\}$, we substitute $\gamma^{(r)}_{ik}(u)$ and $\lambda^{(r)}_{ik}(u)$ for $z_{ik}$ and $v_{ik}$, respectively, in $\ell^1_c\{\boldsymbol{\Theta}(u)\}$
\begin{eqnarray}
Q\{\boldsymbol{\Theta}(u)|\boldsymbol{\Theta}^{(r-1)}(u)\}=\sum_{i=1}^n\sum_{k=1}^K\gamma^{(r)}_{ik}(u)\log\bigg[\pi_k(u)\big(\hat{\alpha}_k\mathcal{N}\{y|m_k(u),\sigma^{2}(u)\})^{\lambda_{ik}^{(r)}(u)}\nonumber\\
    \times((1-\hat{\alpha}_k)\mathcal{N}\{y|m_k(u),\hat{\eta}_k\sigma^2(u)\}\big)^{1-\lambda_{ik}^{(r)}(u)}\bigg]W_h(x_i-u)\label{exp_loglik1}
\end{eqnarray}
Note that if we separately maximize each local log-likelihood $\ell^1_c\{\boldsymbol{\Theta}(u)\}$, for each $u\in \mathcal{U}$, the estimated non-parametric functions $\{\widehat{\boldsymbol{\Theta}}(x_i): i=1,2,\dots,n\}$ may be subject to label-switching. This is because $\gamma^{(r)}_{ik}(u)$ and $\lambda^{(r)}_{ik}(u)$ are not guaranteed to match across all the local points $\mathcal{U}$. The modified EM algorithm replaces the local E-steps with a single global E-step. That is, each $\gamma^{(r)}_{ik}(u)$ and $\lambda^{(r)}_{ik}(u)$ are replaced by $\gamma^{(r)}_{ik}$ and $\lambda^{(r)}_{ik}$, respectively, for all $u\in\mathcal{U}$.  More specifically, at the $r^{th}$ iteration, given $\boldsymbol{\Theta}^{(r-1)}_k=(\boldsymbol{\pi}^{(r-1)}_k,\boldsymbol{\theta}^{(r-1)}_k)$, with $\boldsymbol{\theta}^{(r-1)}_k=(\hat{\alpha}_k,\hat{\mathbf{m}}_k^{(r-1)},\hat{\boldsymbol{\sigma}}^{2(r-1)}_k,\hat{\eta}_k)$, for $k=1,2,\dots,K$, we calculate $\gamma^{(r)}_{ik}$ and $\lambda^{(r)}_{ik}$, respectively, as 
\begin{eqnarray}
    \gamma^{(r)}_{ik}=\frac{\pi_k^{(r-1)}(x_i)f(y_i|x_i;\boldsymbol{\theta}^{(r-1)}_k)}{p(y_i|x_i;\boldsymbol{\Theta}^{(r-1)})}\label{resp1}\\
    \lambda^{(r)}_{ik}=\frac{\hat{\alpha}_k\mathcal{N}\{y_i|m^{(r-1)}_k(x_i),\sigma^2_k(x_i)\}}{f_{CG}(y_i|x_i;\boldsymbol{\theta}^{(r-1)}_k)}\label{resp2}
\end{eqnarray}
Note that \eqref{resp1} and \eqref{resp2} are independent of the local points $\mathcal{U}$. At the M-step, we update $\boldsymbol{\Theta}(u)$, for $u\in \mathcal{U}$, by maximizing 
$Q\{\boldsymbol{\Theta}(u)|\boldsymbol{\Theta}^{(r-1)}(u)\}$ \eqref{exp_loglik1}, which can be written as
\begin{equation}
    Q\{\boldsymbol{\Theta}(u)|\boldsymbol{\Theta}^{(r-1)}(u)\}=Q\{\boldsymbol{\pi}(u)|\boldsymbol{\pi}^{(r-1)}(u)\}+Q\{\mathbf{m}(u),\boldsymbol{\sigma}^2(u)|\mathbf{m}^{(r-1)}(u),\boldsymbol{\sigma}^{2(r-1)}(u)\}
\end{equation}
where
\begin{eqnarray}
Q\{\boldsymbol{\pi}(u)|\boldsymbol{\pi}^{(r-1)}(u)\}&=&\sum_{i=1}^n\sum_{k=1}^K\gamma^{(r)}_{ik}\log(\pi_k(u))W_h(x_i-u),
\end{eqnarray}
\begin{eqnarray}
Q\{\mathbf{m}(u),\boldsymbol{\sigma}^2(u)|\mathbf{m}^{(r-1)}(u),\boldsymbol{\sigma}^{2(r-1)}(u)\}=\sum_{i=1}^n\sum_{k=1}^K\gamma^{(r)}_{ik}\big[ \lambda^{(r)}_{ik}\log\mathcal{N}\{y|m_k(u),\sigma_k^{2}(u)\})+\nonumber\\({1-\lambda^{(r)}_{ik}})\log\mathcal{N}\{y|m_k(x_i),\hat{\eta}_k\sigma_k^2(u)\}\big)\big]W_h(x_i-u)\nonumber
\\
\end{eqnarray}
with respect to $\boldsymbol{\Theta}(u)$ to obtain
\begin{eqnarray}
    \pi^{(r)}_k(u)&=&\frac{\sum_{i=1}^n\gamma^{(r)}_{ik}W_h(x_i-u)}{\sum_{i=1}^nW_h(x_i-u)}\label{mixprop1}\\
    m^{(r)}_k(u)&=&\frac{\sum_{i=1}^nw^{(r)}_{ik}(u)W_h(x_i-u)y_i}{\sum_{i=1}^nw^{(r)}_{ik}(u)W_h(x_i-u)}\label{regfun1}\\
    \sigma^{2(r)}_k(u)&=&\frac{\sum_{i=1}^nw^{(r)}_{ik}(u)W_h(x_i-u)[y_i-m^{(r)}_k(u)]^2}{\sum_{i=1}^n\gamma^{(r)}_{ik}W_h(x_i-u)}\label{varfun1}
\end{eqnarray}
where
\begin{equation}
    w_{ik}^{(r)}(u)=\gamma^{(r)}_{ik}\bigg(\lambda^{(r)}_{ik}+\frac{1-\lambda^{(r)}_{ik}}{\hat{\eta}_k}\bigg)\quad\text{for }i=1,2,\dots,n\text{ and }k=1,2,\dots,K
\end{equation}
Note that $\gamma^{(r)}_{ik}$ and $\lambda^{(r)}_{ik}$ in \eqref{mixprop1}-\eqref{varfun1} are kept constant across all the local points $\mathcal{U}$. This avoids the label-switching problem mentioned above.\\
We repeat the above E- and M-step until convergence. To obtain $\{\hat{\boldsymbol{\Theta}}(x_i): i=1,2,\dots,n\}$, the estimated non-parametric functions, for $x_i\notin \mathcal{U}$, we interpolate over $\{\hat{\boldsymbol{\Theta}}(u):u\in\mathcal{U}\}$.\\
Given $\{\hat{\boldsymbol{\Theta}}(x_i): i=1,2,\dots,n\}$, we maximize $ \ell^2\{\boldsymbol{\alpha},\boldsymbol{\eta}\}$ using the classical EM algorithm. We first introduce two latent variables $\mathbf{z}_i$ and $\mathbf{v}_i$, for $i=1,2\dots,n$, defined as above. The complete-data are $\{(x_i,y_i,\mathbf{z}_i,\mathbf{v}_i):i=1,2,\dots,n\}$ and the complete-data log-likelihood is 
\begin{eqnarray}
    \ell^2_c\{\boldsymbol{\alpha},\boldsymbol{\eta}\}=\sum_{i=1}^n\sum_{k=1}^Kz_{ik}\log\bigg[\widehat{\pi}_k(x_i)\big(\alpha_k\mathcal{N}\{y|\widehat{m}_k(x_i),\widehat{\sigma}^{2}(x_i)\})^{v_{ik}}\nonumber\\
    \times((1-\alpha_k)\mathcal{N}\{y|\widehat{m}_k(x_i),\eta_k\widehat{\sigma}^2(u)\}\big)^{1-v_{ik}}\bigg]\label{comp_loglik2}
\end{eqnarray}
which can be written as 
\begin{equation}
\ell^2_c\{\boldsymbol{\alpha},\boldsymbol{\eta}\}=\ell^2_{1c}\{\boldsymbol{\alpha}\}+\ell^2_{2c}\{\boldsymbol{\eta}\}+\text{const}
\end{equation}
where
\begin{eqnarray}
\ell^2_{1c}\{\boldsymbol{\alpha}\}&=&\sum_{i=1}^n\sum_{k=1}^Kz_{ik}\big[v_{ik}\log\alpha_k+(1-v_{ik})\log(1-\alpha_k)\big]\\
\ell^2_{2c}\{\boldsymbol{\eta}\}&=&\sum_{i=1}^n\sum_{k=1}^Kz_{ik}(1-v_{ik})\log\mathcal{N}\{y_i|\hat{m}_k(x_i),\eta_k\hat{\sigma}^2_k(x_i)\}
\end{eqnarray}
and "const" represents all the terms that are independent of both $\boldsymbol{\alpha}$ and $\boldsymbol{\eta}$.\\
In the E-step, at the $r^{th}$ iteration, given $\boldsymbol{\Theta}^{(r-1)}_k=(\hat{\boldsymbol{\pi}}_k,\boldsymbol{\theta}^{(r-1)}_k)$, with $\boldsymbol{\theta}^{(r-1)}_k=(\alpha^{(r-1)}_k,\hat{\mathbf{m}}_k,\hat{\boldsymbol{\sigma}}^{2}_k,\eta^{(r-1)}_k)$, for $k=1,2,\dots,K$, we calculate the conditional expectations $\mathbb{E}\big[z_{ik}|x_i,y_i,\boldsymbol{\alpha}^{(r-1)},\boldsymbol{\eta}^{(r-1)}\big]$ and  $\mathbb{E}\big[v_{ik}|x_i,y_i,\mathbf{z}_i,\boldsymbol{\alpha}^{(r-1)},\boldsymbol{\eta}^{(r-1)}\big]$ as follows, respectively,
\begin{eqnarray}
\gamma^{1(r)}_{ik}=\frac{\widehat{\pi}_k(x_i)f(y_i|x_i;\boldsymbol{\theta}^{(r-1)}_k)}{p(y_i|x_i;\boldsymbol{\Theta}^{(r-1)})}\label{resp3}\\
\lambda^{1(r)}_{ik}=\frac{\alpha^{(r-1)}_k\mathcal{N}\{y_i|\widehat{m}_k(x_i),\widehat{\sigma}^2_k(x_i)\}}{f_{CG}(y_i|x_i;\boldsymbol{\theta}^{(r-1)}_k)}\label{resp4}
\end{eqnarray}
In the M-step, we update $\boldsymbol{\alpha}$ and $\boldsymbol{\eta}$, respectively, by maximizing the conditional expectation of $\ell^2_c\{\boldsymbol{\alpha},\boldsymbol{\eta}\}$
\begin{eqnarray}
    Q\{\boldsymbol{\alpha},\boldsymbol{\eta}|\boldsymbol{\alpha}^{(r-1)},\boldsymbol{\eta}^{(r-1)}\}=\sum_{i=1}^n\sum_{k=1}^K\gamma^{1(r)}_{ik}\log\bigg[\widehat{\pi}_k(x_i)\big(\alpha_k\mathcal{N}\{y|\widehat{m}_k(x_i),\widehat{\sigma}^{2}(x_i)\})^{\lambda^{1(r)}_{ik}}\nonumber\\
    \times((1-\alpha_k)\mathcal{N}\{y|\widehat{m}_k(x_i),\eta_k\widehat{\sigma}^2(u)\}\big)^{1-\lambda^{1(r)}_{ik}}\bigg]\label{exp_loglik2}
\end{eqnarray}
with respect to $\boldsymbol{\alpha}$ and $\boldsymbol{\eta}$, respectively, as
\begin{eqnarray}
    \alpha_k^{(r)}=\frac{1}{n^{(r)}_k}\sum_{i=1}^n\gamma^{1(r)}_{ik}\lambda^{1(r)}_{ik}\label{alpha1}\\
    \eta^{(r)}_k=\max\bigg\{1,\frac{b^{(r)}_k}{a^{(r)}_k}\bigg\}\label{eta1}
\end{eqnarray}
where
\begin{eqnarray}
    n^{(r)}_k&=&\sum_{i=1}^n\gamma^{1(r)}_{ik}\nonumber\\
    a^{(r)}_k&=&\sum_{i=1}^n\gamma^{1(r)}_{ik}\big(1-\lambda^{1(r)}_{ik}\big)\nonumber\\
    b^{(r)}_k&=&\sum_{i=1}^n\gamma^{1(r)}_{ik}\big(1-\lambda^{1(r)}_{ik}\big)\big(y_i-\widehat{m}_k(x_i)\big)^2/\widehat{\sigma}_k^{2(r-1)}(x_i)\nonumber~\text{ for } k=1,2,\dots,K
\end{eqnarray}
The update equation for $\eta_k$ in \eqref{eta1} follows because of the constraint $\eta_k>1$, for $k=1,2,\dots,K$.\\
We repeat the above E- and M-steps until convergence to obtain the updated parameter estimates.\\
The above backfitting EM-type algorithm is summarised in Algorithm \ref{algo1}.
\begin{algorithm}
\caption{Backfitting estimation procedure via the EM algorithm for model \eqref{model4}}\label{algo1}
\begin{algorithmic}[1]
\STATE \textbf{Step 0:} Obtain the initial parameter estimates $\{(\hat{\alpha}_k,\hat{\eta}_k):k=1,2,\dots,K\}$ by fitting a CGMLRs model.
\STATE \textbf{Step 1:} Given $\{(\widehat{\alpha}_k,\widehat{\eta}_k):k=1,2,\dots,K\}$, estimate $\{\boldsymbol{\Theta}(u):u\in\mathcal{U}\}$ by maximizing $\ell^1$. Let $\{\boldsymbol{\Theta}^{(0)}(u):u\in\mathcal{U}\}$ be the initial estimate of $\{\boldsymbol{\Theta}(u): u\in\mathcal{U}\}$. At the $r^{th}$ iteration,
\begin{enumerate}
\item[] \textbf{E-step:} we calculate the posterior probabilities $\gamma^{(r)}_{ik}$ and $\lambda^{(r)}_{ik}$, for $i=1,2,\dots,n$ and $k=1,2,\dots,K$, using \eqref{resp1} and \eqref{resp2}, respectively.
\item[] \textbf{M-step:} we update $\{\boldsymbol{\Theta}(u):u\in\mathcal{U}\}$ using \eqref{mixprop1}-\eqref{varfun1}.
\item[] Interpolate over $\{\boldsymbol{\Theta}^{(r)}(u):u\in\mathcal{U}\}$ to obtain $\{\boldsymbol{\Theta}^{(r)}(x_i):i=1,2,\dots,n\}$, for $x_i\notin\mathcal{U}$.
\item[] Repeat the above E- and M-step until convergence.
\end{enumerate}
\STATE \textbf{Step 2:} Given $\{\widehat{\boldsymbol{\Theta}}(x_i):i=1,2,\dots,n\}$ from \textbf{Step 1}, improve the parameter estimates $\{(\widehat{\alpha}_k,\widehat{\eta}_k):k=1,2,\dots,K\}$ by maximizing $\ell^2$. Let $\alpha^{(0)}_k=\widehat{\alpha}_k$ and $\eta^{(0)}_k=\widehat{\eta}_k$, for $k=1,2,\dots,K$, be the initial parameter estimates. At the $r^{th}$ iteration,
\begin{enumerate}
    \item [] \textbf{E-step:} calculate $\gamma^{1(r)}_{ik}$ and $\lambda^{1(r)}_{ik}$, for $i=1,2,\dots,n$ and $k=1,2,\dots,K$, using \eqref{resp3} and \eqref{resp4}, respectively.
    \item [] \textbf{M-step:} we update $\boldsymbol{\alpha}$ and $\boldsymbol{\eta}$ using \eqref{alpha1} and \eqref{eta1}, respectively.
    \item [] Repeat the above E- and M-step until convergence. 
\end{enumerate}
\end{algorithmic}
\end{algorithm}
Since \textbf{Step 2} and \textbf{Step 3} of Algorithm \ref{algo1} are performed only once, the algorithm is a one-step backfitting algorithm. To improve the model estimates, Algorithm \ref{algo1} can be implemented by iterating between \textbf{Step 2} and \textbf{Step 3} until convergence. However, this is computationally expensive. The second proposed fitting algorithm is a less computationally intensive approximation of the latter iterative backfitting algorithm. 
\paragraph{Algorithm 2}
We now propose a modified expectation-conditional maximization (ECM) \cite{meng1993} algorithm to simultaneously maximizes the log-likelihoods $\ell^1$ and $\ell^2$. The ECM algorithm is a variant of the classical EM algorithm that splits a complex M-step into simpler conditional maximization (CM-) steps.\\
Note that in order to update $\{\boldsymbol\Theta(u):u\in\mathcal{U}\}$, we need to know $\boldsymbol{\eta}$. On the other hand, in order to update $\boldsymbol{\eta}$, we need to know $\{\boldsymbol\Theta(x_i):i=1,2,\dots,n\}$. To simultaneously update $\{\boldsymbol\Theta(u):u\in\mathcal{U}\}$ and $\boldsymbol{\eta}$, at the $r^{th}$ iteration we split the M-step into two CM-steps. In the first CM-step, CM-step 1, we update $\{\boldsymbol\Theta(u):u\in\mathcal{U}\}$ and $\boldsymbol{\alpha}$ with $\boldsymbol{\eta}$ fixed at $\boldsymbol{\eta}^{(r-1)}$. In the second CM-step, CM-step 2, we update $\boldsymbol{\eta}$ with $\{\boldsymbol\Theta(x_i):i=1,2,\dots,n\}$ fixed at $\{\boldsymbol\Theta^{(r)}(x_i):i=1,2,\dots,n\}$.\\
As before, define the two latent variables $\mathbf{z}_i$ and $\mathbf{v}_i$, for $i=1,2\dots,n$. In the E-step, at the $r^{th}$ iteration, given $\boldsymbol{\Theta}^{(r-1)}_k=(\boldsymbol{\pi}^{(r-1)}_k,\boldsymbol{\theta}^{(r-1)}_k)$, with $\boldsymbol{\theta}^{(r-1)}_k=(\alpha^{(r-1)}_k,\mathbf{m}^{(r-1)}_k,\boldsymbol{\sigma}^{2(r-1)}_k,\eta^{(r-1)}_k)$, for $k=1,2,\dots,K$, we calculate the conditional expectations $\mathbb{E}\big[z_{ik}|x_i,y_i,\boldsymbol{\Theta}^{(r-1)}\big]$ and  $\mathbb{E}\big[v_{ik}|x_i,y_i,\mathbf{z}_i,\boldsymbol{\Theta}^{(r-1)}\big]$ as follows, respectively,
\begin{eqnarray}
\gamma^{2(r)}_{ik}&=&\frac{{\pi}^{(r-1)}_k(x_i)f(y_i|x_i;\boldsymbol{\theta}^{(r-1)}_k)}{p(y_i|x_i;\boldsymbol{\Theta}^{(r-1)})}\label{resp5}\\
\lambda^{2(r)}_{ik}&=&\frac{\alpha^{(r-1)}_k\mathcal{N}\{y_i|m^{(r-1)}_k(x_i),\sigma^{2(r-1)}_k(x_i)\}}{f_{CG}(y_i|x_i;\boldsymbol{\theta}^{(r-1)}_k)}\label{resp6}
\end{eqnarray}
At the first CM step, given $\eta^{(r-1)}_k$, we update $\boldsymbol{\pi}_k$, $\alpha_k$, $\mathbf{m}_k$ and $\boldsymbol{\sigma}^2_k$, respectively, as follows
\begin{eqnarray}
    \pi^{(r)}_k(u)&=&\frac{\sum_{i=1}^n\gamma^{2(r)}_{ik}W_h(x_i-u)}{\sum_{i=1}^nW_h(x_i-u)}\label{mixprop2}\\
    \alpha_k^{(r)}&=&\frac{1}{n^{2(r)}_k}\sum_{i=1}^n\gamma^{2(r)}_{ik}\lambda^{2(r)}_{ik}\label{alpha2}\\
    m^{(r)}_k(u)&=&\frac{\sum_{i=1}^nw^{2(r)}_{ik}W_h(x_i-u)y_i}{\sum_{i=1}^nw^{2(r)}_{ik}W_h(x_i-u)}\label{regfun2}\\
    \sigma^{2(r)}_k(u)&=&\frac{\sum_{i=1}^nw^{2(r)}_{ik}W_h(x_i-u)[y_i-m^{(r)}_k(u)]^2}{\sum_{i=1}^n\gamma^{2(r)}_{ik}W_h(x_i-u)}\label{varfun2}
\end{eqnarray}
where
\begin{eqnarray}
    n^{2(r)}_k&=&\sum_{i=1}^n\gamma^{2(r)}_{ik}\nonumber\\
    w_{ik}^{2(r)}(u)&=&\gamma^{2(r)}_{ik}\bigg(\lambda^{2(r)}_{ik}+\frac{1-\lambda^{2(r)}_{ik}}{\eta^{(r-1)}_k}\bigg)\quad\text{for }i=1,2,\dots,n\text{ and }k=1,2,\dots,K\nonumber
\end{eqnarray}
At the second CM-step, given $\boldsymbol{\pi}^{2(r)}_k$, $\alpha^{(r)}_k$, $\mathbf{m}^{(r)}_k$ and $\boldsymbol{\sigma}^{2(r)}_k$, update $\eta_k$ as follows
\begin{eqnarray}
    \eta^{(r)}_k=\max\bigg\{1,\frac{b^{2(r)}_k}{a^{2(r)}_k}\bigg\}\label{eta2}
\end{eqnarray}
where
\begin{eqnarray}
    a^{2(r)}_k&=&\sum_{i=1}^n\gamma^{2(r)}_{ik}\big(1-\lambda^{2(r)}_{ik}\big)\nonumber\\
    b^{2(r)}_k&=&\sum_{i=1}^n\gamma^{2(r)}_{ik}\big(1-\lambda^{2(r)}_{ik}\big)\big(y_i-m^{(r)}_k(x_i)\big)^2/\sigma_k^{2(r)}(x_i)\nonumber
\end{eqnarray}
for $i=1,2,\dots,n$ and $k=1,2,\dots,K$.\\
We repeat the above E- and CM-steps until convergence. The above algorithm is summarised in Algorithm \ref{algo2}.
\begin{algorithm}
\caption{Global estimation procedure via a modified ECM algorithm for model \eqref{model4}}\label{algo2}
\begin{algorithmic}[1]
\STATE \textbf{Step 0:} Estimate a $K-$component CGMLRs model to obtain the estimates $(\widehat{\pi}_k,\widehat{\alpha}_k,\widehat{\boldsymbol{\beta}}_k,\widehat{\sigma}^2_k,\widehat{\eta}_k)$, for $k=1,2,\dots,K$. Let $\boldsymbol{\Theta}^{(0)}_k=(\boldsymbol{\pi}^{(0)}_k,\boldsymbol{\theta}^{(0)}_k)$, for $k=1,2,\dots,K$, where $\boldsymbol{\theta}^{(0)}_k=(\alpha^{(0)}_k,\mathbf{m}^{(0)}_k,\boldsymbol{\sigma}^{2(0)}_k,\eta^{(0)}_k)$, with $\boldsymbol{\pi}^{(0)}_k=\{\widehat{\pi}_k:i=1,2,\dots,n\}$, $\mathbf{m}^{(0)}_k=\{\mathbf{x}_i^\top\widehat{\boldsymbol{\beta}}_k:i=1,2,\dots,n\}$ and $\boldsymbol{\sigma}^{2(0)}_k=\{\hat{\sigma}^2_k:i=1,2,\dots,n\}$.
\STATE \textbf{Step 1:} Let $\boldsymbol{\Theta}^{(r-1)}$ be the estimate of $\boldsymbol{\Theta}$ at the $(r-1)^{th}$ iteration. At the $r^{th}$ iteration,
\begin{enumerate}
\item[] \textbf{E-step:} we calculate the posterior probabilities $\gamma^{2(r)}_{ik}$ and $\lambda^{2(r)}_{ik}$, for $i=1,2,\dots,n$ and $k=1,2,\dots,K$, using \eqref{resp5} and \eqref{resp6}, respectively.
\item[] \textbf{CM-step 1:} Given $\eta^{(r-1)}_k$, we update $\boldsymbol{\pi}_k$, $\alpha_k$, $\mathbf{m}_k$ and $\boldsymbol{\sigma}^2_k$ using \eqref{mixprop2}-\eqref{varfun2}, for $k=1,2,\dots,K$. Interpolate over $\{(\boldsymbol{\pi}^{(r)}(u),\mathbf{m}^{(r)}(u),\boldsymbol{\sigma}^2(u)):u\in\mathcal{U}\}$ to obtain $\{(\boldsymbol{\pi}^{(r)}(x_i),\mathbf{m}^{(r)}(x_i),\boldsymbol{\sigma}^2(x_i)):i=1,2,\dots,n\}$, for $x_i\notin\mathcal{U}$.
\item[] \textbf{CM-step 2:} Given $\boldsymbol{\pi}^{(r)}_k$, $\alpha^{(r)}_k$, $\mathbf{m}^{(r)}_k$ and $\boldsymbol{\sigma}^{2(r)}_k$, we update $\eta_k$ using \eqref{eta2}, for $k=1,2,\dots,K$.
\item [] Repeat the above E- and CM-steps until convergence.
\end{enumerate}
\end{algorithmic}
\end{algorithm}
\section{Semi-parametric contaminated Gaussian mixtures of non-parametric regressions (SPCGMRs)}\label{SPCGMRs}
In this section, we define the proposed SPCGMRs model and derive the proposed estimation procedure for the model.
\subsection{Model definition}
For some problems, the NPCGMRs model introduced in section \ref{NPCGMRs} may be too general and unnecessarily complex. In such cases, a parsimonius model may be enough to explain the data. The SPCGMRs model assumes that the mixing proportions and the variances $\pi_k$ and $\sigma^2_k$, respectively, for $k=1,2,\dots,K$, are parametric as in the CGMRs model, whereas the component regression functions $m_k(x)$, for $k=1,2,\dots,K$, are assumed to be non-parametric, as in the NPCGMRs model \eqref{model4}. Thus, the SPCGMRs model is a natural extension of the CGMRs and the NPCGMRs model.\\
Let $Z$ be a component indicator with a discrete distribution $P(Z=k)=\pi_k$, for $k=1,2,\dots,K$, where each $\pi_k>0$ and $\sum_{k=1}^K\pi_k=1$. Conditional on $Z=k$ and $X=x$, $y$ is assumed to follow a contaminated Gaussian distribution with density function
\begin{equation}
    f_{CG}(y|X=x;\theta_k)=\alpha_k\mathcal{N}\{y|m_k(x),\sigma^{2}_k\}+(1-\alpha_k)\mathcal{N}\{y|m_k(x),\eta_k\sigma_k^2\}
\end{equation}
Given only $X=x$, $y$ is said to follow a mixture of contaminated Gaussian distributions with density function
\begin{align}
p(y|X=x;\boldsymbol{\Theta})&=\sum_{k}^K\pi_kf_{CG}(y|X=x;\theta_k)\nonumber\\
    &=\sum_{k=1}^K\pi_k\big[\alpha_k\mathcal{N}\{y|m_k(x),\sigma_k^{2}\}+(1-\alpha_k)\mathcal{N}\{y|m_k(x),\eta_k\sigma_k^2\}\big]\label{model5}
\end{align}
Since each $m_k(x)$ is non-parametric, model \eqref{model5} is a semi-parametric contaminated Gaussian mixture of non-parametric regressions (SPCGMRs). The identifiability of the SPCGMRs model follows from an extension of proposition 1 of \cite{xiang2018} and the fact that the parametric mixture of contaminated Gaussian distributions is identifiable as shown by \cite{punzo2016}.
\subsection{Model estimation}
Consider a random sample $\{(x_i,y_i):i=1,2,\dots,n\}$ from \eqref{model5}. The corresponding log-likelihood function is
\begin{equation}
    \ell\{\boldsymbol{\Theta}\}=\sum_{i=1}^n\log\bigg[\sum_{k=1}^K\pi_k\big(\alpha_k\mathcal{N}\{y_i|m_k(x_i),\sigma_k^{2}\}+(1-\alpha_k)\mathcal{N}\{y_i|m_k(x_i),\eta_k\sigma_k^2\}\big)\bigg]\label{loglik5}
\end{equation}
where $\boldsymbol{\Theta}=\{\boldsymbol{\alpha},\boldsymbol{\eta},\boldsymbol{\pi},\boldsymbol{\sigma}^2,\mathbf{m}\}$, $\boldsymbol{\alpha}=\{\alpha_1,\dots,\alpha_K\}$, $\boldsymbol{\eta}=\{\eta_1,\dots,\eta_K\}$, $\boldsymbol{\pi}=\{\pi_1,\dots,\pi_K\}$, $\boldsymbol{\sigma}^2=\{\sigma^2_1,\dots,\sigma^2_K\}$ and $\mathbf{m}=\{\mathbf{m}_1,\dots,\mathbf{m}_K\}$, with $\mathbf{m}_k=\{m_k(x_i):i=1,2,\dots,n\}$.\\
Note that $\boldsymbol{\Theta}$ is made up of parametric $\{\boldsymbol{\alpha},\boldsymbol{\eta},\boldsymbol{\pi},\boldsymbol{\sigma}^2\}$ and non-parametric terms $\mathbf{m}$. Thus, the log-likelihood function cannot be directly maximized. As in section \eqref{backfit1}, we propose a backfitting estimation procedure to alternately estimate the parametric terms and the non-parametric terms. Given estimates of the parametric terms, denoted by $(\hat{\boldsymbol{\alpha}},\hat{\boldsymbol{\eta}},\hat{\boldsymbol{\pi}},\hat{\boldsymbol{\sigma}}^2)$, we estimate the non-parametric term $\mathbf{m}$ by maximizing the following locally weighted log-likelihood, over all $u\in\mathcal{U}$,
\begin{eqnarray}
    \ell^1\{\mathbf{m}(u)\}=\sum_{i=1}^n\log\bigg[\sum_{k=1}^K\hat{\pi}_k\big(\hat{\alpha}_k\mathcal{N}\{y_i|m_k(u),\hat{\sigma}^{2}\}+\nonumber\\
    (1-\hat{\alpha}_k)\mathcal{N}\{y_i|m_k(u),\hat{\eta}_k\hat{\sigma}^2\}\big)\bigg]W_h(x_i-u)\label{loglik6}
\end{eqnarray}
where $\mathbf{m}(u)=\{m_k(u):k=1,2,\dots,K\}$ and $\mathcal{U}$ is the set of local points on the domain of the covariate $x$. Let $\hat{\mathbf{m}}(u)$ be the estimate of $\mathbf{m}(u)$, for $u\in \mathcal{U}$. To obtain $\hat{\mathbf{m}}(x_i)$, for $x_i\notin \mathcal{U}$, we interpolate over $\hat{\mathbf{m}}(u)$, for $u\in \mathcal{U}$.\\
Given $\{\hat{\mathbf{m}}(x_i):i=1,2,\dots,n\}$, we update the estimates of the parametric terms $(\hat{\boldsymbol{\alpha}},\hat{\boldsymbol{\eta}},\hat{\boldsymbol{\pi}},\hat{\boldsymbol{\sigma}}^2)$ by maximizing 
\begin{eqnarray}
    \ell^2\{\boldsymbol{\alpha},\boldsymbol{\eta},\boldsymbol{\pi},\boldsymbol{\sigma}^2\}=\sum_{i=1}^n\log\bigg[\sum_{k=1}^K\pi_k\big(\alpha_k\mathcal{N}\{y_i|\widehat{m}_k(x_i),\sigma^{2}\}+\nonumber\\
    (1-\alpha_k)\mathcal{N}\{y_i|\widehat{m}_k(x_i),\eta_k\sigma^2\}\big)\bigg]W_h(x_i-u)\label{loglik6}
\end{eqnarray}
Let $(\tilde{\boldsymbol{\alpha}},\tilde{\boldsymbol{\eta}},\tilde{\boldsymbol{\pi}},\tilde{\boldsymbol{\sigma}}^2)$ be the parameter estimates obtained from maximizing the log-likelihood function \eqref{loglik6}. The above estimation procedure can be implemented using a special case of either Algorithm \ref{algo1} or Algorithm \ref{algo2}. Since Algorithm \ref{algo2} is less computationally intensive, we propose to use its special case for obtaining the estimates of model \eqref{model5} by simultaneously maximizing the log-likelihood functions $\ell^1\{\mathbf{m}(u)\}$ and $\ell^2\{\boldsymbol{\alpha},\boldsymbol{\eta},\boldsymbol{\pi},\boldsymbol{\sigma}^2\}$. 
\paragraph{Algorithm 3}
We now propose a modified ECM algorithm to simultaneously maximize the log-likelihoods $\ell^1\{\mathbf{m}(u)\}$ and $\ell^2\{\boldsymbol{\alpha},\boldsymbol{\eta},\boldsymbol{\pi},\boldsymbol{\sigma}^2\}$. Note that in order to update $\{\mathbf{m}(u):u\in\mathcal{U}\}$ and $\boldsymbol{\sigma}^2$, we need to know $\boldsymbol{\eta}$. On the other hand, in order to update $\boldsymbol{\eta}$, we need to know $\{\mathbf{m}(x_i):i=1,2,\dots,n\}$ and $\boldsymbol{\sigma}^2$. In order to simultaneously update $\{\mathbf{m}(u):u\in\mathcal{U}\}$, $\boldsymbol{\sigma}^2$ and $\boldsymbol{\eta}$, at the $r^{th}$ iteration, we split the M-step into two conditional-maximization (CM)-steps. In the first CM-step, CM-step 1, we update $\{\mathbf{m}(u):u\in\mathcal{U}\}$ and $\boldsymbol{\sigma}^2$, with $\boldsymbol{\eta}$ fixed at $\boldsymbol{\eta}^{(r-1)}$. In the second CM-step, CM-step 2, we update $\boldsymbol{\eta}$ with $\{\mathbf{m}(x_i):i=1,2,\dots,n\}$ and $\boldsymbol{\sigma}^2$ fixed at $\{\mathbf{m}^{(r)}(x_i):i=1,2,\dots,n\}$ and $\boldsymbol{\sigma}^{2(r)}$.\\
As before, define the two latent variables $\mathbf{z}_i$ and $\mathbf{v}_i$, for $i=1,2\dots,n$. In the E-step, at the $r^{th}$ iteration, given $\boldsymbol{\Theta}^{(r-1)}_k=(\pi^{(r-1)}_k,\boldsymbol{\theta}^{(r-1)}_k)$, with $\boldsymbol{\theta}^{(r-1)}_k=(\alpha^{(r-1)}_k,\mathbf{m}^{(r-1)}_k,\sigma^{2(r-1)}_k,\eta^{(r-1)}_k)$, for $k=1,2,\dots,K$, we calculate the conditional expectations $\mathbb{E}\big[z_{ik}|x_i,y_i,\boldsymbol{\Theta}^{(r-1)}\big]$ and  $\mathbb{E}\big[v_{ik}|x_i,y_i,\mathbf{z}_i,\boldsymbol{\Theta}^{(r-1)}\big]$ as follows, respectively,
\begin{eqnarray}
\gamma^{2(r)}_{ik}&=&\frac{{\pi}^{(r-1)}_kf(y_i|x_i;\boldsymbol{\theta}^{(r-1)}_k)}{p(y_i|x_i;\boldsymbol{\Theta}^{(r-1)})}\label{resp7}\\
\lambda^{2(r)}_{ik}&=&\frac{\alpha^{(r-1)}_k\mathcal{N}\{y_i|m^{(r-1)}_k(x_i),\sigma^{2(r-1)}_k\}}{f_{CG}(y_i|x_i;\boldsymbol{\theta}^{(r-1)}_k)}\label{resp8}
\end{eqnarray}
At the first CM step, given $\eta^{(r-1)}_k$, we update $\boldsymbol{\pi}_k$, $\alpha_k$, $\mathbf{m}_k$ and $\sigma^2_k$, respectively, as follows
\begin{eqnarray}
    \pi^{(r)}_k&=&\frac{\sum_{i=1}^n\gamma^{2(r)}_{ik}}{n}\label{mixprop3}\\
    \alpha_k^{(r)}&=&\frac{1}{n^{2(r)}_k}\sum_{i=1}^n\gamma^{2(r)}_{ik}\lambda^{2(r)}_{ik}\label{alpha3}\\
    m^{(r)}_k(u)&=&\frac{\sum_{i=1}^nw^{2(r)}_{ik}W_h(x_i-u)y_i}{\sum_{i=1}^nw^{2(r)}_{ik}W_h(x_i-u)}\quad\text{for }u\in\mathcal{U}\label{regfun3}\\
    \sigma^{2(r)}_k&=&\frac{\sum_{i=1}^nw^{2(r)}_{ik}[y_i-m^{(r)}_k(x_i)]^2}{n^{2(r)}_k}\label{varfun3}
\end{eqnarray}
where
\begin{eqnarray}
    n^{2(r)}_k&=&\sum_{i=1}^n\gamma^{2(r)}_{ik}\nonumber\\
    w_{ik}^{2(r)}&=&\gamma^{2(r)}_{ik}\bigg(\lambda^{2(r)}_{ik}+\frac{1-\lambda^{2(r)}_{ik}}{\eta^{(r-1)}_k}\bigg)\quad\text{for }i=1,2,\dots,n\text{ and }k=1,2,\dots,K\nonumber
\end{eqnarray}
Note: The $\{m^{(r)}_k(x_i):i=1,2,\dots,n;k=1,2,\dots,K\}$ in \eqref{varfun3}, are obtained by interpolating over $m^{(r)}_k(u)$, for $u\in\mathcal{U}$ and $k=1,2,\dots,K$.\\
At the second CM-step, given $\pi^{2(r)}_k$, $\alpha^{(r)}_k$, $\mathbf{m}^{(r)}_k$ and $\sigma^{2(r)}_k$, update $\eta_k$ as follows
\begin{eqnarray}
    \eta^{(r)}_k=\max\bigg\{1,\frac{b^{2(r)}_k}{a^{2(r)}_k}\bigg\}\label{eta3}
\end{eqnarray}
where
\begin{eqnarray}
    a^{2(r)}_k&=&\sum_{i=1}^n\gamma^{2(r)}_{ik}\big(1-\lambda^{2(r)}_{ik}\big)\nonumber\\
    b^{2(r)}_k&=&\sum_{i=1}^n\gamma^{2(r)}_{ik}\big(1-\lambda^{2(r)}_{ik}\big)\big(y_i-m^{(r)}_k(x_i)\big)^2/\sigma_k^{2(r)}\nonumber
\end{eqnarray}
for $i=1,2,\dots,n$ and $k=1,2,\dots,K$.\\
We repeat the above E- and CM-steps until convergence. The above algorithm is summarised in Algorithm \ref{algo3}.
\begin{algorithm}
\caption{Global estimation procedure via a modified ECM algorithm for model \eqref{model5}}\label{algo3}
\begin{algorithmic}[1]
\STATE \textbf{Step 0:} Estimate a $K-$component CGMLRs model to obtain the estimates $(\widehat{\pi}_k,\widehat{\alpha}_k,\widehat{\boldsymbol{\beta}}_k,\widehat{\sigma}^2_k,\widehat{\eta}_k)$, for $k=1,2,\dots,K$. Let $\boldsymbol{\Theta}^{(0)}_k=(\pi^{(0)}_k,\boldsymbol{\theta}^{(0)}_k)$, for $k=1,2,\dots,K$, where $\boldsymbol{\theta}^{(0)}_k=(\alpha^{(0)}_k,\mathbf{m}^{(0)}_k,\sigma^{2(0)}_k,\eta^{(0)}_k)$, with $\mathbf{m}^{(0)}_k=\{\mathbf{x}_i^\top\widehat{\boldsymbol{\beta}}_k:i=1,2,\dots,n\}$.
\STATE \textbf{Step 1:} Let $\boldsymbol{\Theta}^{(r-1)}$ be the estimate of $\boldsymbol{\Theta}$ at the $(r-1)^{th}$ iteration. At the $r^{th}$ iteration,
\begin{enumerate}
\item[] \textbf{E-step:} we calculate the posterior probabilities $\gamma^{2(r)}_{ik}$ and $\lambda^{2(r)}_{ik}$, for $i=1,2,\dots,n$ and $k=1,2,\dots,K$, using \eqref{resp7} and \eqref{resp8}, respectively.
\item[] \textbf{CM-step 1:} Given $\eta^{(r-1)}_k$, we update $\pi_k$, $\alpha_k$, $\mathbf{m}_k$ and $\sigma^2_k$ using \eqref{mixprop3}-\eqref{varfun3}, for $k=1,2,\dots,K$. Interpolate over $\{\mathbf{m}^{(r)}(u):u\in\mathcal{U}\}$ to obtain $\{\mathbf{m}^{(r)}(x_i):i=1,2,\dots,n\}$, for $x_i\notin\mathcal{U}$.
\item[] \textbf{CM-step 2:} Given $\pi^{(r)}_k$, $\alpha^{(r)}_k$, $\mathbf{m}^{(r)}_k$ and $\sigma^{2(r)}_k$, we update $\eta_k$ using \eqref{eta3}, for $k=1,2,\dots,K$.
\item [] Repeat the above E- and CM-steps until convergence.
\end{enumerate}
\end{algorithmic}
\end{algorithm}
\section{Model-based clustering}\label{model-based}
In this section, we discuss two important benefits of modelling data with the proposed models: (1) model-based clustering of observations into various groups or components and (2) model-based detection of outliers. The first benefit is common to all mixture models whereas the second benefit is unique to contaminated Gaussian mixture models. For the proposed models: NPCGMRs \eqref{model4} and SPCGMRs \eqref{model5}, clustering a data point $(x_i,y_i)$ involves two steps. In the first step, we assign $(x_i,y_i)$ to a given component using the maximum \emph{a posteriori} (MAP) operator
\begin{equation}\label{MAP}
\text{MAP}(\gamma_{ik})=\begin{cases}
    1\quad\text{if $\max_j\gamma_{ij}$ occurs in component $k=j$}\\
    0\quad\text{otherwise}
\end{cases}
\end{equation}
where $\gamma_{ik}$ is the posterior probability calculated at the E-step of the EM or ECM algorithm, using \eqref{resp1} and \eqref{resp3} or \eqref{resp5} and \eqref{resp7}, respectively.\\
Using \eqref{MAP}, the data point $(x_i,y_i)$ will be assigned to the $j^{th}$ component. In the second step, we establish whether or not the data point $(x_i,y_i)$ is an outlier in the $j^{th}$ component. For this purpose, we consider $\lambda_{ij}$, the posterior probability calculated at the E-step of the EM or ECM algorithm using \eqref{resp2} and \eqref{resp4} or \eqref{resp6} and \eqref{resp8}, respectively. As in \cite{mazza2020}, if $\lambda_{ij}<0.5$, then the data point $(x_i,y_i)$ can be classified as an outlier. 
\section{Model selection}\label{model-select}
For the proposed models, model selection involves selecting $K$, the number of components and $h$, the bandwidth or smoothing parameter used in local-likelihood estimation to estimate the non-parametric functions. The optimal value of $(K,h)$ can be obtained using a data-driven criteria such as the Akaike information criteria (AIC) \citep{akaike1974}
\begin{equation}
    \text{AIC}=-2\ell+2\times df,
\end{equation}
the Bayesian information criteria (BIC) \citep{schwarz1978}
\begin{equation}
    \text{BIC}=-2\ell+df\times\log{(n)}
\end{equation}
or the integrated classification likelihood (ICL) \citep{biernacki2000}
\begin{equation}
    \text{ICL}=-2\ell_c+df\times\log{(n)}
\end{equation}
where $\ell$ and $\ell_c$ is the maximum log-likelihood value and maximum complete-data log-likelihood value, respectively, at convergence of the EM or ECM and $df$ is the degrees of freedom ($df$) which measure the complexity of the fitted model. For a parametric mixture model, $df$ is obtained by counting the number of parameters in the model. This is, however, not the case for the proposed models because they have both parametric and non-parametric terms. Therefore, the degrees of freedom is a function of the model complexity due to the parameters and the non-parametric functions. As in \cite{huang2013} and \cite{wu2017}, we measure the model complexity due to a non-parametric function using the effective degrees of freedom ($edf$) of a one-dimensional non-parametric function defined as
\begin{equation}
    edf=\tau_Wh^{-1}|\Phi|\bigg\{W(0)-\frac{1}{2}\int W^2(t)dt\bigg\}
\end{equation}
where $\Phi$ is the support of the covariate $x$ in \eqref{model4} or \eqref{model5} and 
\begin{equation}
    \tau_W=\frac{W(0)-\frac{1}{2}\int W^2(t)dt}{\int\{W(t)-\frac{1}{2}W* W(t)\}^2dt}
\end{equation}
The degrees of freedom for model \eqref{model4} and model \eqref{model5} is given by $df=(3K-1)\times edf+2\times K$ and $df=K\times edf+(4K-1)$, respectively. The optimal value of $(K,h)$ is chosen by minimising the AIC, BIC or ICL. We suggest the following strategy for simultaneously choosing $K$ and $h$: Calculate the information criteria over a $2-$dimensional grid with a wide range of values for $(K,h)$ and choose the values $(K,h)$ that minimise the information criteria.  
\section{Simulation study}\label{simulations}
In this section, we conduct intensive Monte Carlo simulations to investigate the behaviour of the proposed models and the finite sample performance of their corresponding proposed estimation procedures. The main aim of the simulations is to demonstrate the robustness of the proposed models when they are fitted to data that exhibits serious departures from the Gaussian assumption. Throughout the simulations, we will compare the proposed models with the non-parametric Gaussian mixture of regressions (NPGMRs; model \eqref{model2}) and the semi-parametric Gaussian mixture of regressions (SPGMRs; model \eqref{model3}). There is currently no statistical software that can be used to estimate these models directly. Therefore, the estimation procedures of these models will be implemented in the \texttt{R} programming language. We make use of the root of the average squared errors (RASE) to measure the performance of the fitted non-parametric functions and parameters, respectively,
    \begin{equation}
        \text{RASE}^2(\mathbf{f})=\frac{1}{n}\sum_{i=1}^n\sum_{k=1}^K\big[\hat{f}_k(x_i)-f_k(x_i)\big]^2
    \end{equation}
    where $\mathbf{f}=\{f_k(x_i):k=1,2,\dots,K;i=1,2,\dots,n\}$, and $f_k(\cdot)$ and $\hat{f}_k(\cdot)$ are the true and estimated functions, respectively.
    \begin{equation}
        \text{RASE}^2(\boldsymbol{\theta})=\frac{1}{K}\sum_{k=1}^K(\hat{\theta}_k-\theta_k)^2
    \end{equation}
    where $\boldsymbol{\theta}=(\theta_1,\theta_2,\dots,\theta_K)$, and $\theta_k$ and $\hat{\theta}_k$ are the true and estimated parameter values, respectively.
\subsection{Numerical experiments}
We now present results from three numerical experiments conducted to: (1) demonstrate the performance of the proposed backfitting estimation procedures (summarized in Algorithm \ref{algo1} and \ref{algo2}) and (2) show the robustness of the proposed models fit to data from various distributions ranging from Gaussian (that is, without outliers) to heavy-tailed (severely contaminated with outliers). In all of our experiments, we generated $500$ samples of sizes $n=500$ and $1000$. Moreover, we considered a single covariate $x$ generated from the uniform distribution on the interval $(0,1)$. As previously pointed out by many authors  (\cite{yao2014}, \cite{mazza2020}, \cite{yu2020}), mixture models suffer from the well-known label switching problem (\cite{stephens2000}, \cite{yao2012}, \cite{skhosana2024}) when evaluating the finite sample performance of the parameters. Since there is no generally accepted approach to address the label-switching problem, we follow the approach of \cite{mazza2020} and \cite{yu2020} by choosing the label that minimizes the distance between the estimated and the true parameter values.
\subsubsection{Experiment 1: Fitting the non-parametric contaminated Gaussian mixture of regressions}
In this experiment, we generate data from a $K=2$ component NPCGMRs model \eqref{model4}. Each $y_i$, for $i=1,2,\dots,n$, is independently generated as follows
\begin{eqnarray}
    y_i=m_k(x_i)+\epsilon_{ik}~~\text{ if } z_i=k~~\text{for }k=1,2\label{dgp1}
\end{eqnarray}
where $m_k(x_i)$ is the regression function of the $k^{th}$ component, $z_i$ is the component indicator of the data point $(x_i,y_i)$, with $P(z_i=k)=\pi_k(x_i)$, for $k=1,2$, and $\epsilon_{ik}\sim \alpha_k\mathcal{N}(0,\sigma_k^2(x_i))+(1-\alpha_k)\mathcal{N}(0,\eta_k\sigma_k^2(x_i))$ is the error term of the $k^{th}$ regression model. The following parametric and non-parametric setting will be used
\begin{eqnarray}    \alpha_1=0.9,\eta_1=20\quad&\text{and}&\quad\alpha_2=0.9,\eta_2=40\nonumber\\
    \pi_1(x)=0.1+0.8\sin(\pi x)\quad&\text{and}&\quad\pi_2(x)=1-\pi_2(x)\nonumber\\
    \sigma_1(x)=0.6\exp{(0.5x)}\quad&\text{and}&\quad \sigma_2(x)=0.5\exp{(-0.2x)}\nonumber\\
    m_1(x)=\text{cos}(3\pi x)\quad&\text{and}&\quad m_2(x)=3-\text{sin}(2\pi x)\nonumber
\end{eqnarray}
The NPCGMRs model is fitted to the 500 data sets using Algorithm \ref{algo1} and Algorithm \ref{algo2}. For comparison, we also fit the NPGMRs model \eqref{model2}. 
Table \ref{tab1} reports the averages (AVG) and standard deviations (SD) of the RASE over $500$ replications. As expected, the NPCGMRs model performs better than the NPGMRs model. Moreover, the NPCGMRs model fitted using Algorithm \ref{algo1} produces results that are comparable to the NPCGMRs model fitted using Algorithm \ref{algo2}. Therefore, in the following numerical studies, we will make use of Algorithm \ref{algo2}.

\begin{table}[h]
\small
\caption{Averages (AVG) and standard deviations (SD) of the RASE calculated over the $500$ replications for Experiment 1}\label{tab1}
\begin{tabular*}{\textwidth}{@{\extracolsep\fill}lcccc}
\hline
&& \multirow{2}{*}{NPGMRs}&\multicolumn{2}{c}{NPCGMRs}\\
\cline{4-5}
&&&Algorithm 1&Algorithm 2\\
\hline
&&\multicolumn{3}{c}{$n=250$}\\
\hline
\multirow{2}{*}{RASE($\mathbf{m}$)}&AVG&0.8322&0.8035&0.8012\\
&SD&0.9729&1.0179&1.0128\\
\multirow{2}{*}{RASE($\boldsymbol{\pi}$)}&AVG&0.1808&0.1679&0.1708\\
&SD&0.1590&0.1490&0.1514\\
\multirow{2}{*}{RASE($\boldsymbol{\sigma}^2$)}&AVG&0.9917&0.4132&0.4276\\
&STD&1.0688&0.2711&0.3135\\
\hline
&&\multicolumn{3}{c}{$n=500$}\\
\hline
\multirow{2}{*}{RASE($\mathbf{m}$)}&AVG&0.4614&0.3937&0.3766\\
&SD&0.6197&0.6731&0.6656\\
\multirow{2}{*}{RASE($\boldsymbol{\pi}$)}&AVG&0.1115&0.1002&0.1007\\
&SD&0.0959&0.0901&0.1002\\
\multirow{2}{*}{RASE($\boldsymbol{\sigma}^2$)}&AVG&0.9834&0.3285&0.3152\\
&SD&0.8082&0.1722&0.1311\\
\hline
&&\multicolumn{3}{c}{$n=1000$}\\
\hline
\multirow{2}{*}{RASE($\mathbf{m}$)}&AVG&0.3579&0.2253&0.2054\\
&SD&0.3276&0.3745&0.3632\\
\multirow{2}{*}{RASE($\boldsymbol{\pi}$)}&AVG&0.0798&0.0671&0.0645\\
&SD&0.0523&0.0568&0.0592\\
\multirow{2}{*}{RASE($\boldsymbol{\sigma}^2$)}&AVG&1.2264&0.2935&0.2754\\
&SD&0.7654&0.1695&0.0915\\
\hline
&&\multicolumn{3}{c}{$n=2000$}\\
\hline
\multirow{2}{*}{RASE($\mathbf{m}$)}&AVG&0.3508&0.1399&0.1176\\
&SD&0.1416&0.0471&0.0242\\
\multirow{2}{*}{RASE($\boldsymbol{\pi}$)}&AVG&0.0710&0.0484&0.0434\\
&SD&0.0277&0.0105&0.0088\\
\multirow{2}{*}{RASE($\boldsymbol{\sigma}^2$)}&AVG&1.4550&0.2811&0.2665\\
&SD&0.5848&0.1117&0.0458\\
\hline
\end{tabular*}
\end{table}

\subsubsection{Experiment 2: Semi-parametric mixture of regressions}
\begin{figure}[!ht]
\centering
\begin{subfigure}{0.35\textwidth}
\centering
\includegraphics[width=\linewidth]{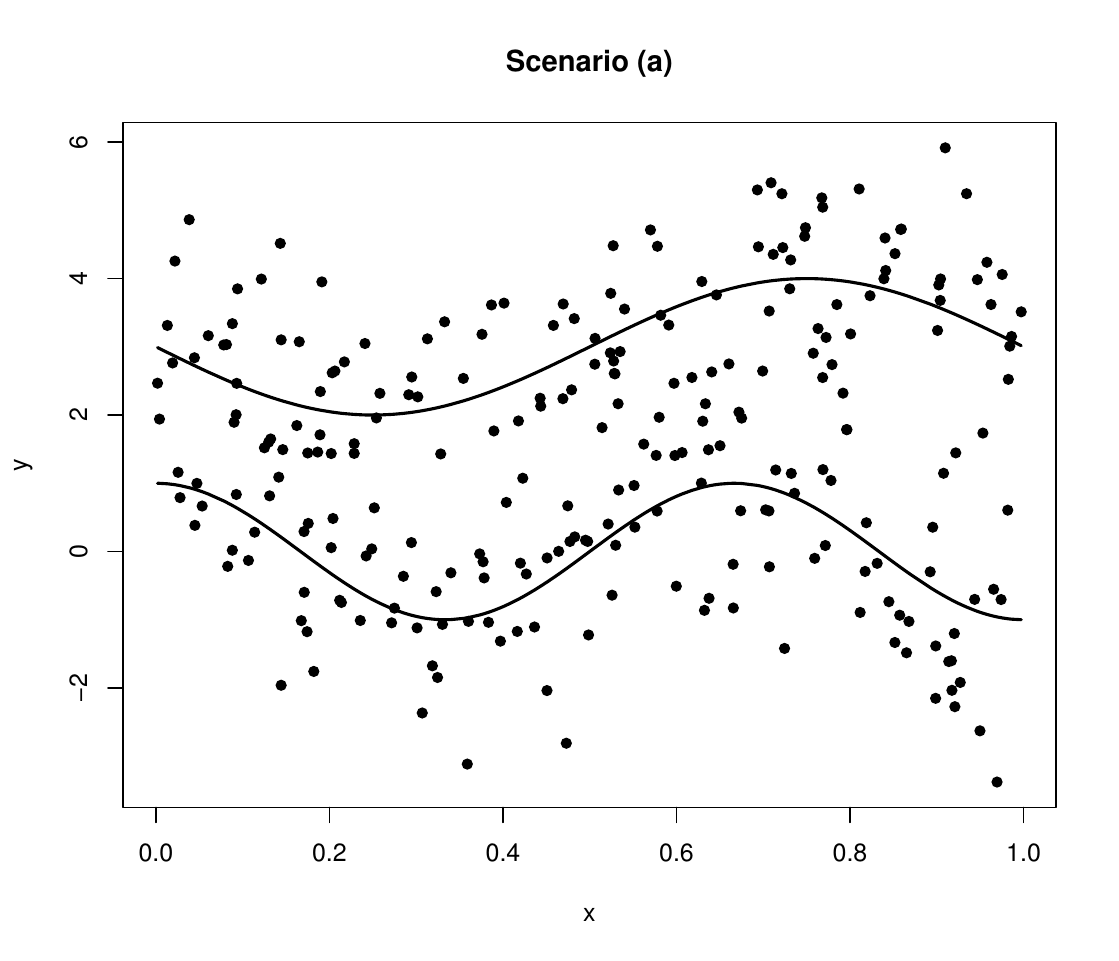}
\label{fig:image1}
\end{subfigure}
\hfill
\begin{subfigure}{0.35\textwidth}
\centering
\includegraphics[width=\linewidth]{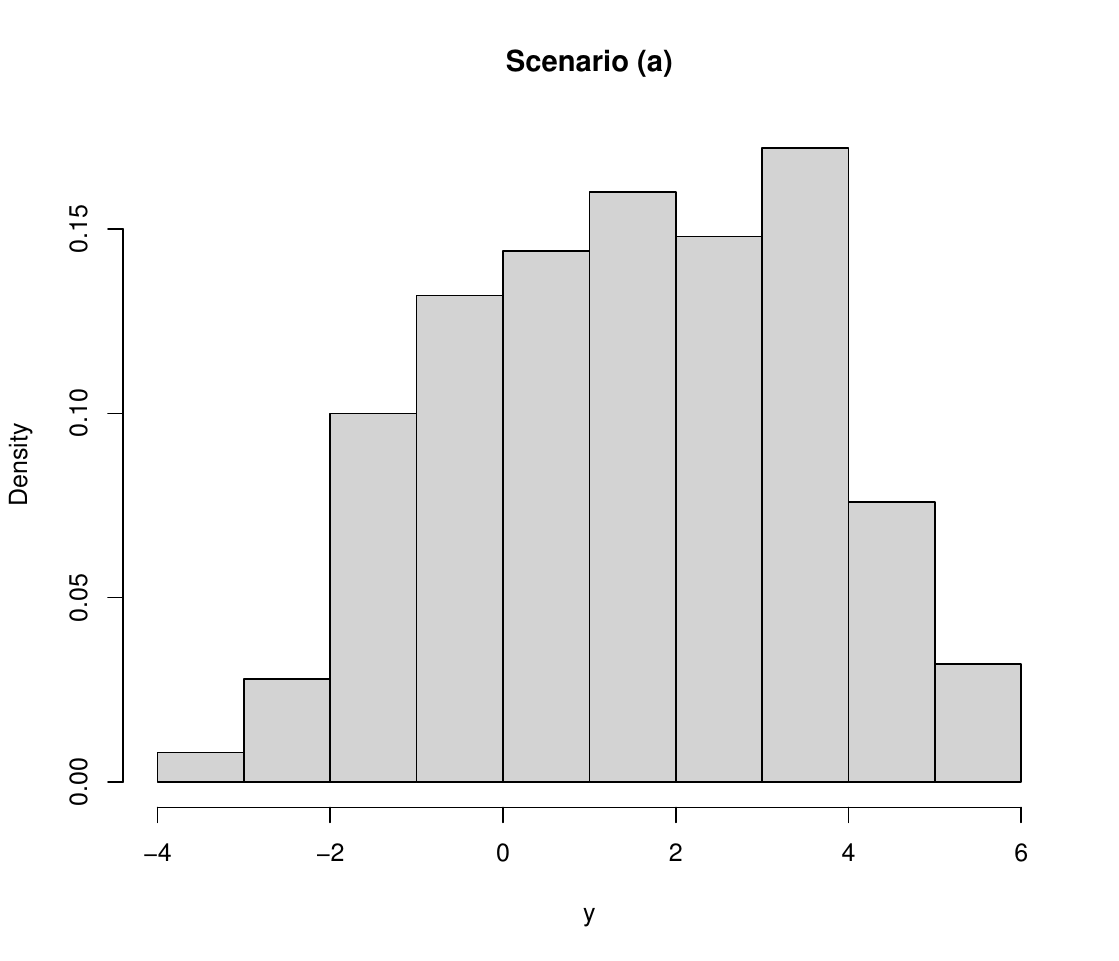}
\label{fig:image2}
\end{subfigure}
\\
\begin{subfigure}{0.35\textwidth}
\centering
\includegraphics[width=\linewidth]{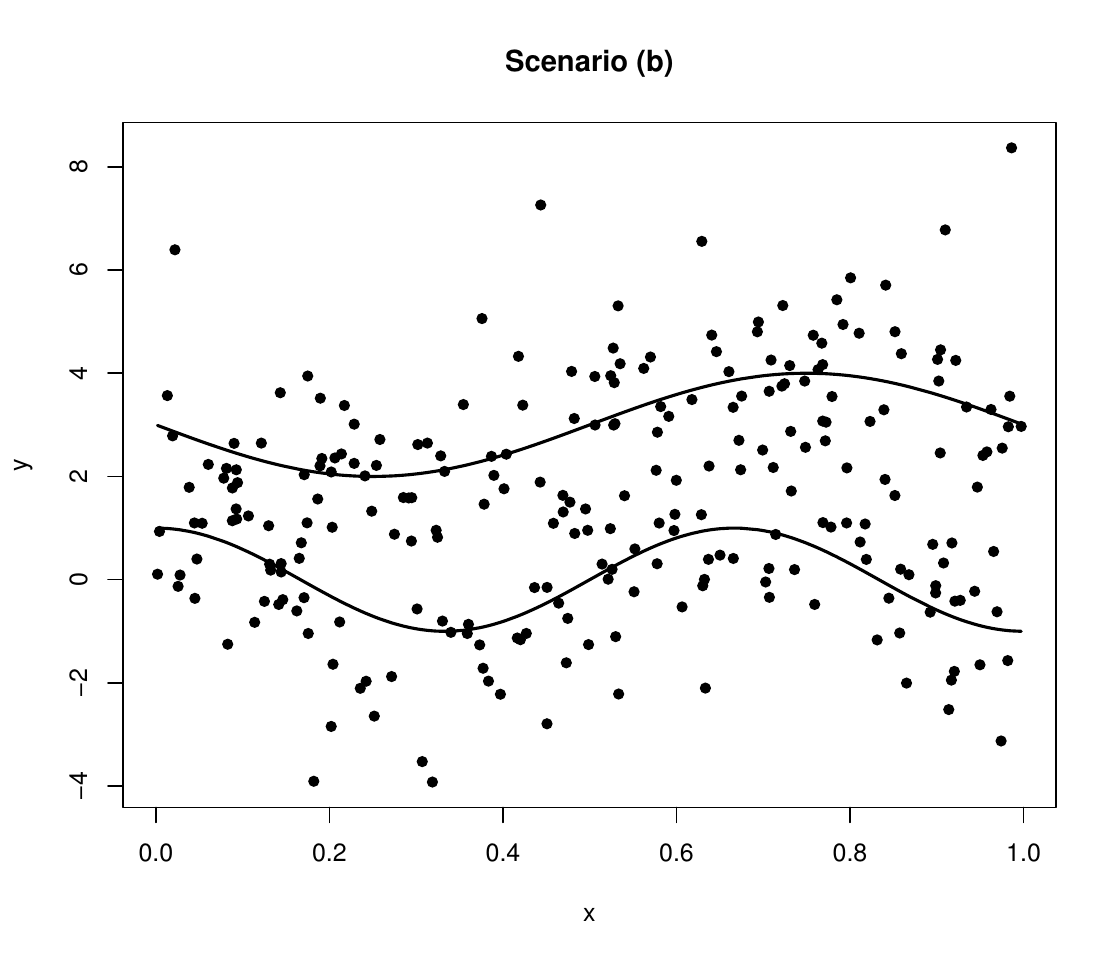}
\label{fig:image2}
\end{subfigure}
\hfill
\begin{subfigure}{0.35\textwidth}
\centering
\includegraphics[width=\linewidth]{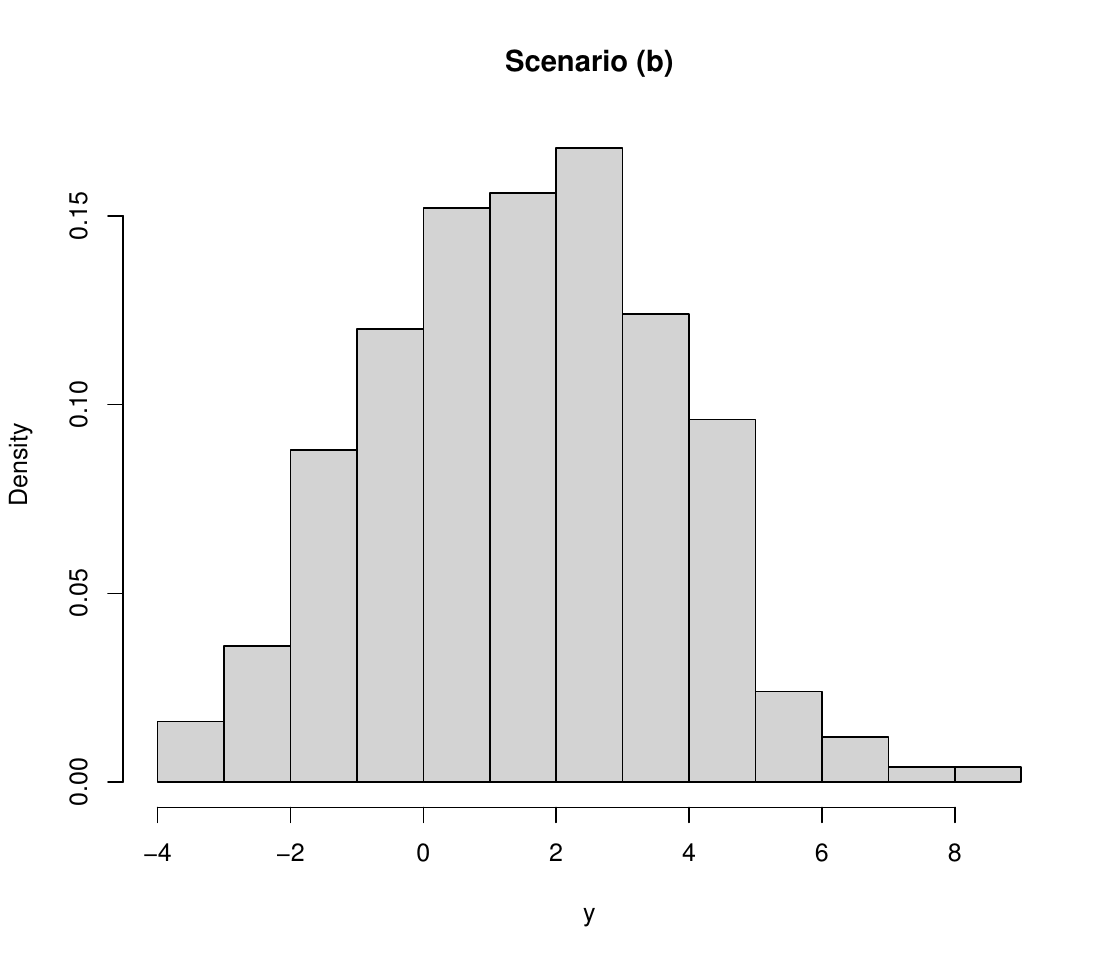}
\label{fig:image1}
\end{subfigure}
\\
\begin{subfigure}{0.35\textwidth}
\centering
\includegraphics[width=\linewidth]{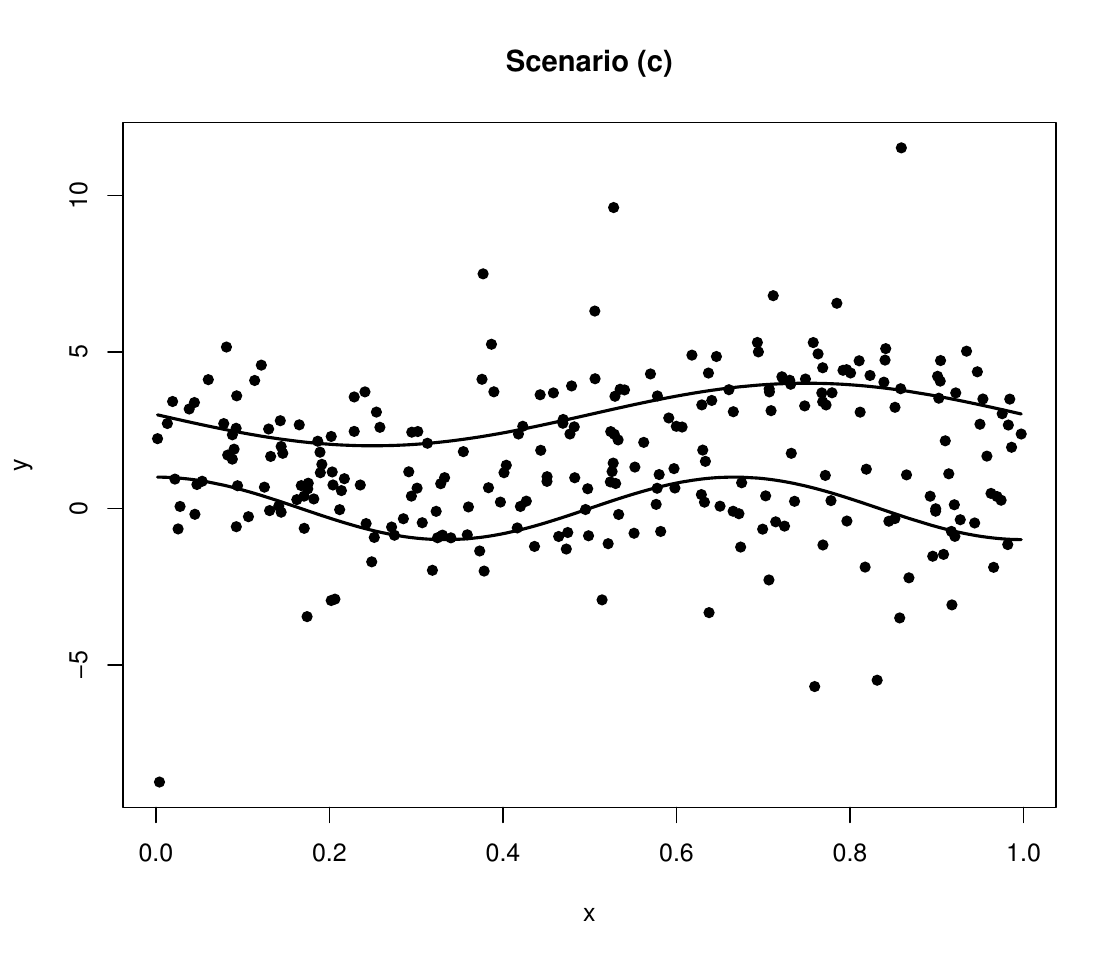}
\label{fig:image1}
\end{subfigure}
\hfill
\begin{subfigure}{0.35\textwidth}
\centering
\includegraphics[width=\linewidth]{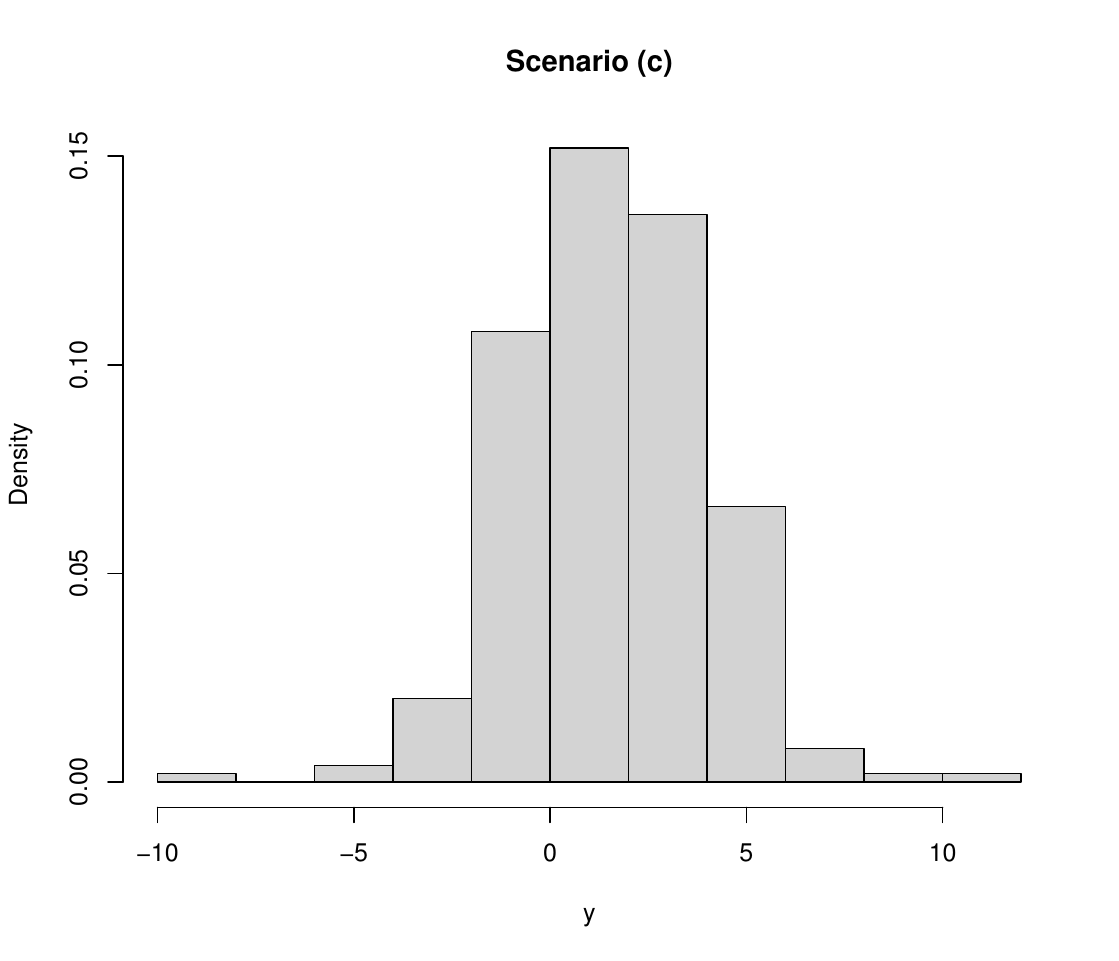}
\label{fig:image2}
\end{subfigure}
\\
\begin{subfigure}{0.35\textwidth}
\centering
\includegraphics[width=\linewidth]{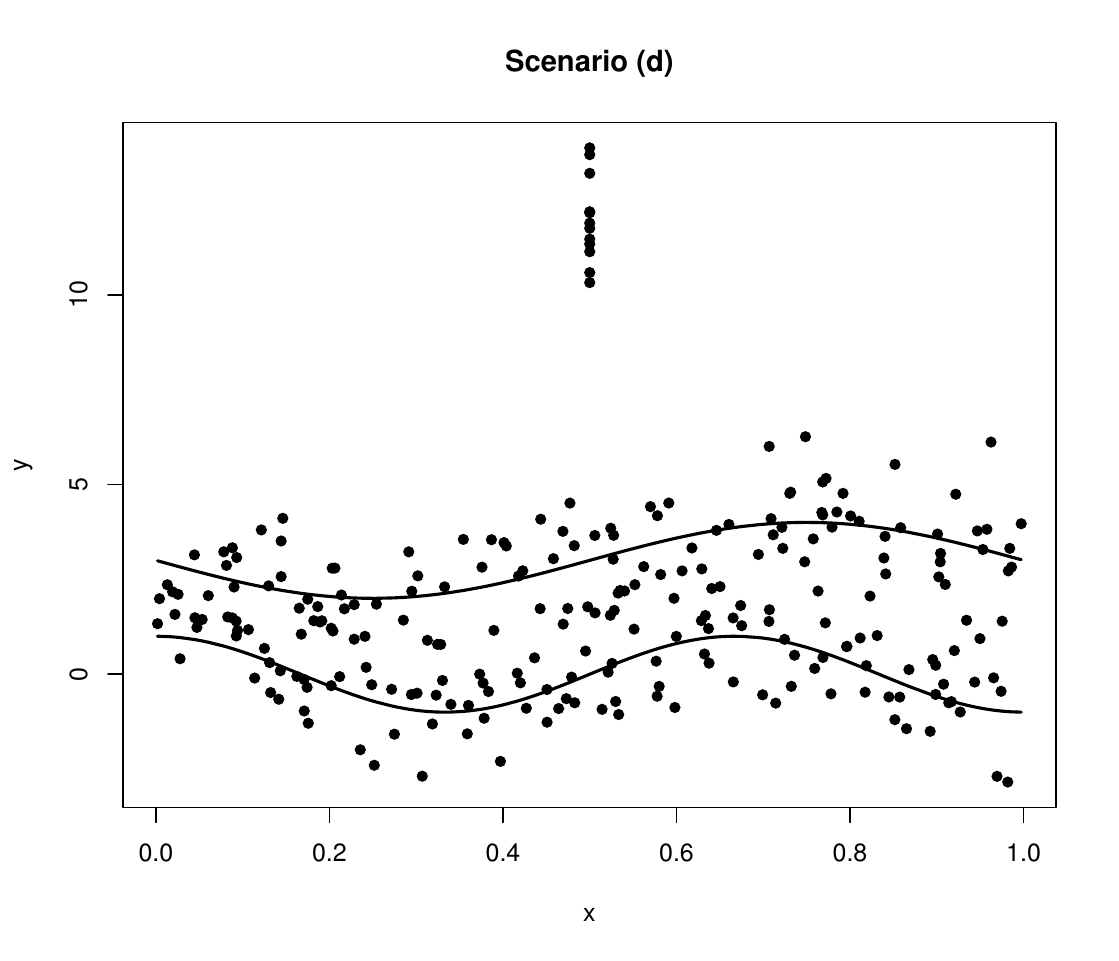}
\label{fig:image1}
\end{subfigure}
\hfill
\begin{subfigure}{0.35\textwidth}
\centering
\includegraphics[width=\linewidth]{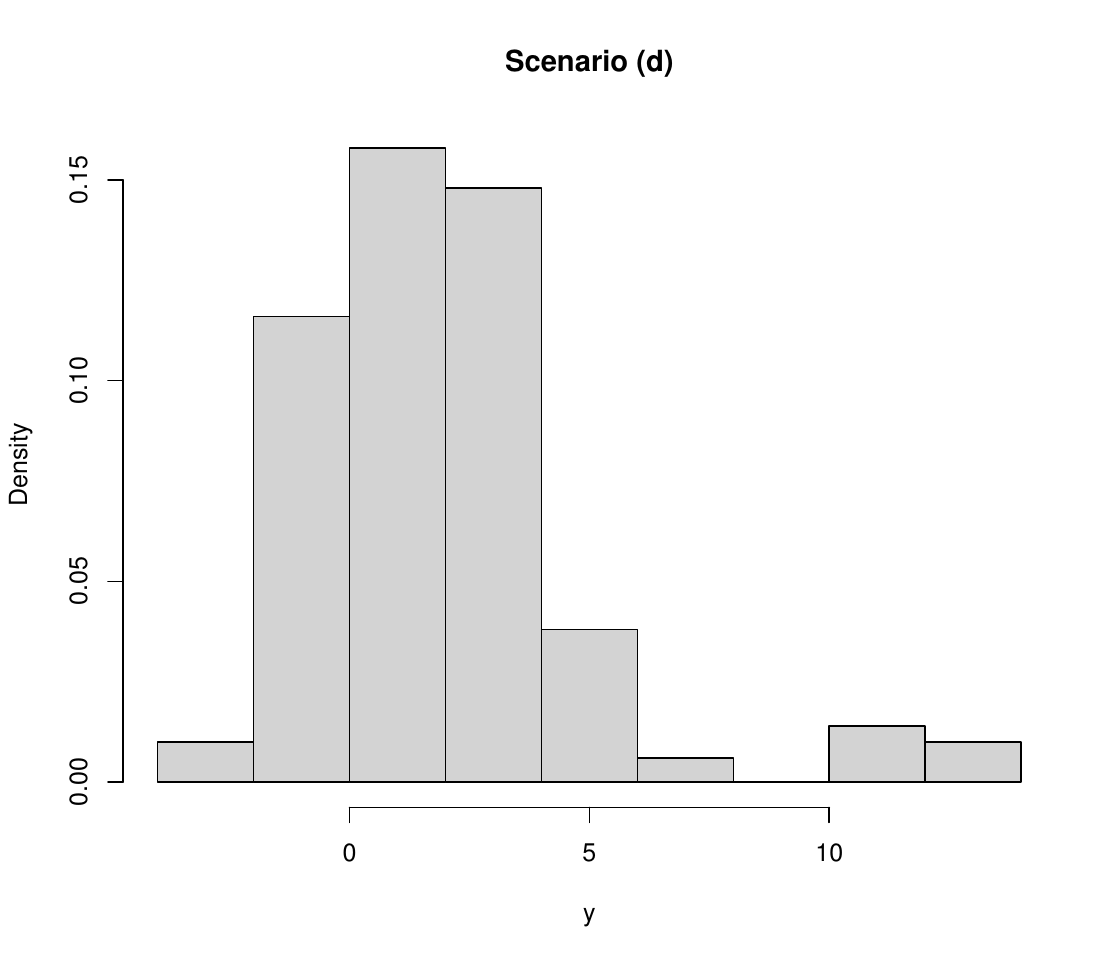}
\label{fig:image2}
\end{subfigure}
\caption{\footnotesize Scatter plots (left-panel) and Histograms (right-panel) of random samples of size $n=250$ generated from each of the Scenarios of Experiment 1. Overlaid on each scatter plot, are the component regression functions.}
\label{fig:scatter_and_hist}
\end{figure}

In this experiment and the next, we assess the robustness of the SPCGMRs and the NPCGMRs models, respectively. The data are generated using the data generating process in \eqref{dgp1}.
where $m_k(x_i)$ is the regression function of the $k^{th}$ component, $z_i$ is the component indicator of the data point $(x_i,y_i)$, with $P(z_i=k)=\pi_k$, for $k=1,2,\dots,K$, and $\epsilon_{ik}$ is the error term of the $k^{th}$ regression model. We consider $K=2$ components. As typically done in the literature (\cite{yao2014}, \cite{yu2020}, \cite{ge2024}), we assume that the component error terms $\epsilon_{k}$, for $k=1,2$, follow the same distribution as $\epsilon$. We consider the following distributions for the error term $\epsilon$:
\begin{enumerate}
    \item[(a)] $\epsilon\sim \mathcal{N}(0,\sigma^2)$
    \item[(b)] $\epsilon\sim 0.9\mathcal{N}(0,\sigma^2)+0.1\mathcal{N}(0,\eta\sigma^2)$, with $\eta=20$
    \item[(c)] $\epsilon\sim \text{t-distribution}(\nu)$, with $\nu=4$ degrees of freedom
    \item[(d)] $\epsilon\sim \mathcal{N}(0,\sigma^2)$ with $5\%$ of the points randomly substituted with high leverage points $(0.5,y_i)$, where $y_i$ is generated from a uniform distribution over the interval $(10,15)$.
    \item[(e)] $\epsilon\sim \mathcal{N}(0,\sigma^2)$ with $5\%$ of the points randomly substituted with noise data points $(x_i,y_i)$, where $x_i$ is generated from a uniform distribution on the interval $(0,1)$ and $y_i$ is generated from a uniform distribution over the interval $(-10,10)$.
\end{enumerate}
All the above scenarios share the following parametric and non-parametric settings:
\begin{eqnarray}
    \pi_1=0.5,\quad \sigma^2=1, \quad m_1(x)=\text{cos}(3\pi x)\quad \text{and}\quad m_2(x)=3-\text{sin}(2\pi x)\nonumber
\end{eqnarray}
The five scenarios above are chosen to represent different situations which may arise when dealing with real-world data. Scenario (a) is used to compare the performance of the proposed methods with traditional methods (with the Gaussian assumption) when the data is actually Gaussian and there are no outliers. Scenarios (b) and (c) considers two heavy-tailed error distributions. Scenario (d) and (e) considers the case of a dataset that is contaminated by outliers and noise data points, respectively.\\
Figure \ref{fig:scatter_and_hist} shows the scatter plots and histograms of typical samples of size $n=250$ from each of the above scenarios. In addition, the true component regression functions are overlaid on the plots. Tables \ref{tab1-1} - \ref{tab1-4} report the averages (AVG) and standard deviations (SD) of the RASE over $500$ replications for all the sample sizes and scenarios. The results can be interpreted as follows. For scenario (a), the SPGMRs model and the SPCGMRs model had a comparable performance and they performed better than the NPGMRs model and the NPCGMRs model. This is to be expected since the data was generated from a mixture of semi-parametric regressions model. However, for scenarios (b) - (e), the SPCGMRs model performed markedly better than all the other models followed by the NPCGMRs model. This highlights the robustness of the latter two models and hence the importance of using a heavy-tailed component distribution when the data are contaminated by outliers. Overall, the SPCGMRs model performed better than all the other models for all the five scenarios considered. Furthermore, the models (SPGMRs and NPGMRs) continued to show poor performance even as we increase the sample size. This further highlights the degree to which a dataset that is contaminated by outliers can influence the model estimates.  
\begin{table}[htbp]
\footnotesize
\caption{Averages (AVG) and standard deviations (SD) of the performance measures calculated over the $500$ replications for $n=250$ in Experiment 2}\label{tab1-1}
\begin{tabular*}{\textwidth}{@{\extracolsep\fill}lccccc}
\hline
&& SPGMRs &NPGMRs&SPCGMRs& NPCGMRs\\
\hline
&&\multicolumn{4}{c}{Scenario (a)}\\
\hline
\multirow{2}{*}{RASE($\mathbf{m}$)}&AVG&0.4171&0.5913&0.4215&0.7985\\
&SD&0.1047&0.1661&0.1198&0.6300\\
\multirow{2}{*}{RASE($\boldsymbol{\pi}$)}&AVG&0.2283&0.2825&0.0769&0.2268\\
&SD&0.1796&0.1154&0.0634&0.0883\\
\multirow{2}{*}{RASE($\boldsymbol{\sigma}^2$)}&AVG&0.2834&0.5942&0.3788&0.6718\\
&SD&0.1390&0.1572&0.2297&0.2925\\
\hline
&&\multicolumn{4}{c}{Scenario (b)}\\
\hline
\multirow{2}{*}{RASE($\mathbf{m}$)}&AVG&0.6238&0.7316&0.5041&0.8911\\
&SD&0.4843&0.3952&0.1646&0.5883\\
\multirow{2}{*}{RASE($\boldsymbol{\pi}$)}&AVG&0.2283&0.2825&0.0769&0.2268\\
&SD&0.1796&0.1154&0.0634&0.0883\\
\multirow{2}{*}{RASE($\boldsymbol{\sigma}^2$)}&AVG&1.0181&1.9608&0.6134&1.3696\\
&SD&0.7911&2.8875&0.3457&0.8672\\
\hline
&&\multicolumn{4}{c}{Scenario (c)}\\
\hline
\multirow{2}{*}{RASE($\mathbf{m}$)}&AVG&1.8068&1.6820&0.4856&0.9383\\
&SD&1.4846&1.1704&0.1402&0.646\\
\multirow{2}{*}{RASE($\boldsymbol{\pi}$)}&AVG&0.2283&0.2825&0.0769&0.2268\\
&SD&0.1796&0.1154&0.0634&0.0883\\
\multirow{2}{*}{RASE($\boldsymbol{\sigma}^2$)}&AVG&6.5092&9.3686&0.4526&1.7292\\
&SD&7.1772&7.1107&0.2648&2.1377\\
\hline
&&\multicolumn{4}{c}{Scenario (d)}\\
\hline
\multirow{2}{*}{RASE($\mathbf{m}$)}&AVG&4.2766&4.2621&0.4673&3.4115\\
&SD&0.6749&0.9658&0.113&2.0354\\
\multirow{2}{*}{RASE($\boldsymbol{\pi}$)}&AVG&0.2283&0.2825&0.0769&0.2268\\
&SD&0.1796&0.1154&0.0634&0.0883\\
\multirow{2}{*}{RASE($\boldsymbol{\sigma}^2$)}&AVG&1.7466&2.2361&0.3449&2.0316\\
&SD&0.8475&2.2972&0.1399&2.1908\\
\hline
&&\multicolumn{4}{c}{Scenario (e)}\\
\hline
\multirow{2}{*}{RASE($\mathbf{m}$)}&AVG&1.3847&1.718&0.4585&0.8655\\
&SD&1.06&1.2418&0.1301&0.6565\\
\multirow{2}{*}{RASE($\boldsymbol{\pi}$)}&AVG&0.2283&0.2825&0.0769&0.2268\\
&SD&0.1796&0.1154&0.0634&0.0883\\
\multirow{2}{*}{RASE($\boldsymbol{\sigma}^2$)}&AVG&4.5631&7.734&0.4047&1.1640\\
&SD&4.5165&5.1331&0.221&1.5004\\
\hline
\end{tabular*}
\end{table}

\begin{table}[htbp]
\footnotesize
\caption{Averages (AVG) and standard deviations (SD) of the performance measures calculated over the $500$ replications for $n=500$ in Experiment 2}\label{tab1-2}
\begin{tabular*}{\textwidth}{@{\extracolsep\fill}lccccc}
\hline
&& SPGMRs &NPGMRs&SPCGMRs& NPCGMRs\\
\hline
&&\multicolumn{4}{c}{Scenario (a)}\\
\hline
\multirow{2}{*}{RASE($\mathbf{m}$)}&AVG&0.3056&0.4686&0.3076&0.6929\\
&SD&0.0662&0.125&0.0671&0.6793\\
\multirow{2}{*}{RASE($\boldsymbol{\pi}$)}&AVG&0.2264&0.2927&0.0519&0.199\\
&SD&0.1855&0.1352&0.0388&0.0874\\
\multirow{2}{*}{RASE($\boldsymbol{\sigma}^2$)}&AVG&0.1917&0.4872&0.2598&0.5462\\
&SD&0.0957&0.1238&0.1641&0.2165\\
\hline
&&\multicolumn{4}{c}{Scenario (b)}\\
\hline
\multirow{2}{*}{RASE($\mathbf{m}$)}&AVG&0.4569&0.613&0.3566&0.7085\\
&SD&0.2135&0.3826&0.0843&0.6059\\
\multirow{2}{*}{RASE($\boldsymbol{\pi}$)}&AVG&0.2264&0.2927&0.0519&0.199\\
&SD&0.1855&0.1352&0.0388&0.0874\\
\multirow{2}{*}{RASE($\boldsymbol{\sigma}^2$)}&AVG&1.0013&1.9345&0.4992&1.0203\\
&SD&0.6619&2.8911&0.2571&0.6237\\
\hline
&&\multicolumn{4}{c}{Scenario (c)}\\
\hline
\multirow{2}{*}{RASE($\mathbf{m}$)}&AVG&1.8341&1.6619&0.3484&0.7025\\
&SD&1.1901&0.8465&0.0802&0.5872\\
\multirow{2}{*}{RASE($\boldsymbol{\pi}$)}&AVG&0.2264&0.2927&0.0519&0.199\\
&SD&0.1855&0.1352&0.0388&0.0874\\
\multirow{2}{*}{RASE($\boldsymbol{\sigma}^2$)}&AVG&9.3419&11.9248&0.2895&0.9053\\
&SD&9.0561&6.4763&0.161&1.4994\\
\hline
&&\multicolumn{4}{c}{Scenario (d)}\\
\hline
\multirow{2}{*}{RASE($\mathbf{m}$)}&AVG&4.1137&4.2283&0.3647&4.5439\\
&SD&0.2547&0.4662&0.0787&1.2283\\
\multirow{2}{*}{RASE($\boldsymbol{\pi}$)}&AVG&0.2264&0.2927&0.0519&0.199\\
&SD&0.1855&0.1352&0.0388&0.0874\\
\multirow{2}{*}{RASE($\boldsymbol{\sigma}^2$)}&AVG&1.5202&1.6309&0.2617&1.8976\\
&SD&0.2889&1.1122&0.1059&1.8712\\
\hline
&&\multicolumn{4}{c}{Scenario (e)}\\
\hline
\multirow{2}{*}{RASE($\mathbf{m}$)}&AVG&1.4532&2.1741&0.3302&0.7061\\
&SD&0.9754&1.3303&0.0745&0.6865\\
\multirow{2}{*}{RASE($\boldsymbol{\pi}$)}&AVG&0.2264&0.2927&0.0519&0.199\\
&SD&0.1855&0.1352&0.0388&0.0874\\
\multirow{2}{*}{RASE($\boldsymbol{\sigma}^2$)}&AVG&5.7157&11.0103&0.2635&0.7372\\
&SD&5.3678&5.2838&0.1271&0.8434\\
\hline
\end{tabular*}
\end{table}

\begin{table}[htbp]
\footnotesize
\caption{Averages (AVG) and standard deviations (SD) of the performance measures calculated over the $500$ replications for $n=1000$ in Experiment 2}\label{tab1-3}
\begin{tabular*}{\textwidth}{@{\extracolsep\fill}lccccc}
\hline
&& SPGMRs &NPGMRs&SPCGMRs& NPCGMRs\\
\hline
&&\multicolumn{4}{c}{Scenario (a)}\\
\hline
\multirow{2}{*}{RASE($\mathbf{m}$)}&AVG&0.2287&0.3823&0.2296&0.547\\
&SD&0.0431&0.0886&0.0425&0.6204\\
\multirow{2}{*}{RASE($\boldsymbol{\pi}$)}&AVG&0.2286&0.3131&0.0387&0.1724\\
&SD&0.1909&0.1626&0.0291&0.0875\\
\multirow{2}{*}{RASE($\boldsymbol{\sigma}^2$)}&AVG&0.1295&0.3746&0.1726&0.4169\\
&SD&0.0641&0.0892&0.1096&0.1578\\
\hline
&&\multicolumn{4}{c}{Scenario (b)}\\
\hline
\multirow{2}{*}{RASE($\mathbf{m}$)}&AVG&0.3793&0.5822&0.2588&0.6069\\
&SD&0.6534&0.545&0.0502&0.6657\\
\multirow{2}{*}{RASE($\boldsymbol{\pi}$)}&AVG&0.2286&0.3131&0.0387&0.1724\\
&SD&0.1909&0.1626&0.0291&0.0875\\
\multirow{2}{*}{RASE($\boldsymbol{\sigma}^2$)}&AVG&0.9918&2.5189&0.365&0.7018\\
&SD&0.5611&5.7127&0.1796&0.4081\\
\hline
&&\multicolumn{4}{c}{Scenario (c)}\\
\hline
\multirow{2}{*}{RASE($\mathbf{m}$)}&AVG&1.849&1.7705&0.2588&0.602\\
&SD&0.7957&0.6732&0.0505&0.6748\\
\multirow{2}{*}{RASE($\boldsymbol{\pi}$)}&AVG&0.2286&0.3131&0.0387&0.1724\\
&SD&0.1909&0.1626&0.0291&0.0875\\
\multirow{2}{*}{RASE($\boldsymbol{\sigma}^2$)}&AVG&11.8983&15.4406&0.1862&0.5613\\
&SD&10.5759&6.3656&0.108&0.5731\\
\hline
&&\multicolumn{4}{c}{Scenario (d)}\\
\hline
\multirow{2}{*}{RASE($\mathbf{m}$)}&AVG&3.9849&4.1644&0.2921&4.7477\\
&SD&0.1601&0.2296&0.0505&0.7283\\
\multirow{2}{*}{RASE($\boldsymbol{\pi}$)}&AVG&0.2286&0.3131&0.0387&0.1724\\
&SD&0.1909&0.1626&0.0291&0.0875\\
\multirow{2}{*}{RASE($\boldsymbol{\sigma}^2$)}&AVG&1.4478&1.46&0.2082&1.6045\\
&SD&0.1837&0.4994&0.0736&1.1237\\
\hline
&&\multicolumn{4}{c}{Scenario (e)}\\
\hline
\multirow{2}{*}{RASE($\mathbf{m}$)}&AVG&1.3449&2.3695&0.2459&0.5971\\
&SD&0.7033&1.149&0.0467&0.7222\\
\multirow{2}{*}{RASE($\boldsymbol{\pi}$)}&AVG&0.2286&0.3131&0.0387&0.1724\\
&SD&0.1909&0.1626&0.0291&0.0875\\
\multirow{2}{*}{RASE($\boldsymbol{\sigma}^2$)}&AVG&5.9571&15.865&0.1819&0.5324\\
&SD&5.5917&5.1634&0.091&1.1644\\
\hline
\end{tabular*}
\end{table}

\begin{table}[htbp]
\footnotesize
\caption{Averages (AVG) and standard deviations (SD) of the performance measures calculated over the $500$ replications for $n=2000$ in Experiment 2}\label{tab1-4}
\begin{tabular*}{\textwidth}{@{\extracolsep\fill}lccccc}
\hline
&& SPGMRs &NPGMRs&SPCGMRs& NPCGMRs\\
\hline
&&\multicolumn{4}{c}{Scenario (a)}\\
\hline
\multirow{2}{*}{RASE($\mathbf{m}$)}&AVG&0.1733&0.3233&0.1738&0.4419\\
&SD&0.0301&0.0633&0.0303&0.5489\\
\multirow{2}{*}{RASE($\boldsymbol{\pi}$)}&AVG&0.2233&0.3174&0.0307&0.1506\\
&SD&0.1963&0.1803&0.0228&0.0795\\
\multirow{2}{*}{RASE($\boldsymbol{\sigma}^2$)}&AVG&0.0909&0.3011&0.112&0.3325\\
&SD&0.0509&0.0546&0.0671&0.124\\
\hline
&&\multicolumn{4}{c}{Scenario (b)}\\
\hline
\multirow{2}{*}{RASE($\mathbf{m}$)}&AVG&0.2632&0.493&0.1949&0.4515\\
&SD&0.1537&0.2343&0.0357&0.5573\\
\multirow{2}{*}{RASE($\boldsymbol{\pi}$)}&AVG&0.2233&0.3174&0.0307&0.1506\\
&SD&0.1963&0.1803&0.0228&0.0795\\
\multirow{2}{*}{RASE($\boldsymbol{\sigma}^2$)}&AVG&0.9877&2.0449&0.2886&0.5043\\
&SD&0.4211&2.5681&0.1575&0.2641\\
\hline
&&\multicolumn{4}{c}{Scenario (c)}\\
\hline
\multirow{2}{*}{RASE($\mathbf{m}$)}&AVG&1.8442&1.7997&0.1951&0.5741\\
&SD&0.6107&0.4047&0.0356&0.7505\\
\multirow{2}{*}{RASE($\boldsymbol{\pi}$)}&AVG&0.2233&0.3174&0.0307&0.1506\\
&SD&0.1963&0.1803&0.0228&0.0795\\
\multirow{2}{*}{RASE($\boldsymbol{\sigma}^2$)}&AVG&13.997&18.1784&0.1274&0.4858\\
&SD&11.2907&6.1016&0.0673&1.1727\\
\hline
&&\multicolumn{4}{c}{Scenario (d)}\\
\hline
\multirow{2}{*}{RASE($\mathbf{m}$)}&AVG&3.8237&4.0043&0.244&4.5705\\
&SD&0.108&0.1099&0.0354&0.5861\\
\multirow{2}{*}{RASE($\boldsymbol{\pi}$)}&AVG&0.2233&0.3174&0.0307&0.1506\\
&SD&0.1963&0.1803&0.0228&0.0795\\
\multirow{2}{*}{RASE($\boldsymbol{\sigma}^2$)}&AVG&1.3644&1.372&0.177&1.4121\\
&SD&0.1314&0.1078&0.0547&0.4556\\
\hline
&&\multicolumn{4}{c}{Scenario (e)}\\
\hline
\multirow{2}{*}{RASE($\mathbf{m}$)}&AVG&1.2369&2.2159&0.1878&0.4649\\
&SD&0.4505&1.0168&0.0345&0.6499\\
\multirow{2}{*}{RASE($\boldsymbol{\pi}$)}&AVG&0.2233&0.3174&0.0307&0.1506\\
&SD&0.1963&0.1803&0.0228&0.0795\\
\multirow{2}{*}{RASE($\boldsymbol{\sigma}^2$)}&AVG&5.6864&18.9729&0.1327&0.3724\\
&SD&5.5132&4.3244&0.0598&0.5924\\
\hline
\end{tabular*}
\end{table}

\section{Application}\label{app}
\begin{figure}[ht]
    \centering
    \includegraphics[width=0.5\linewidth]{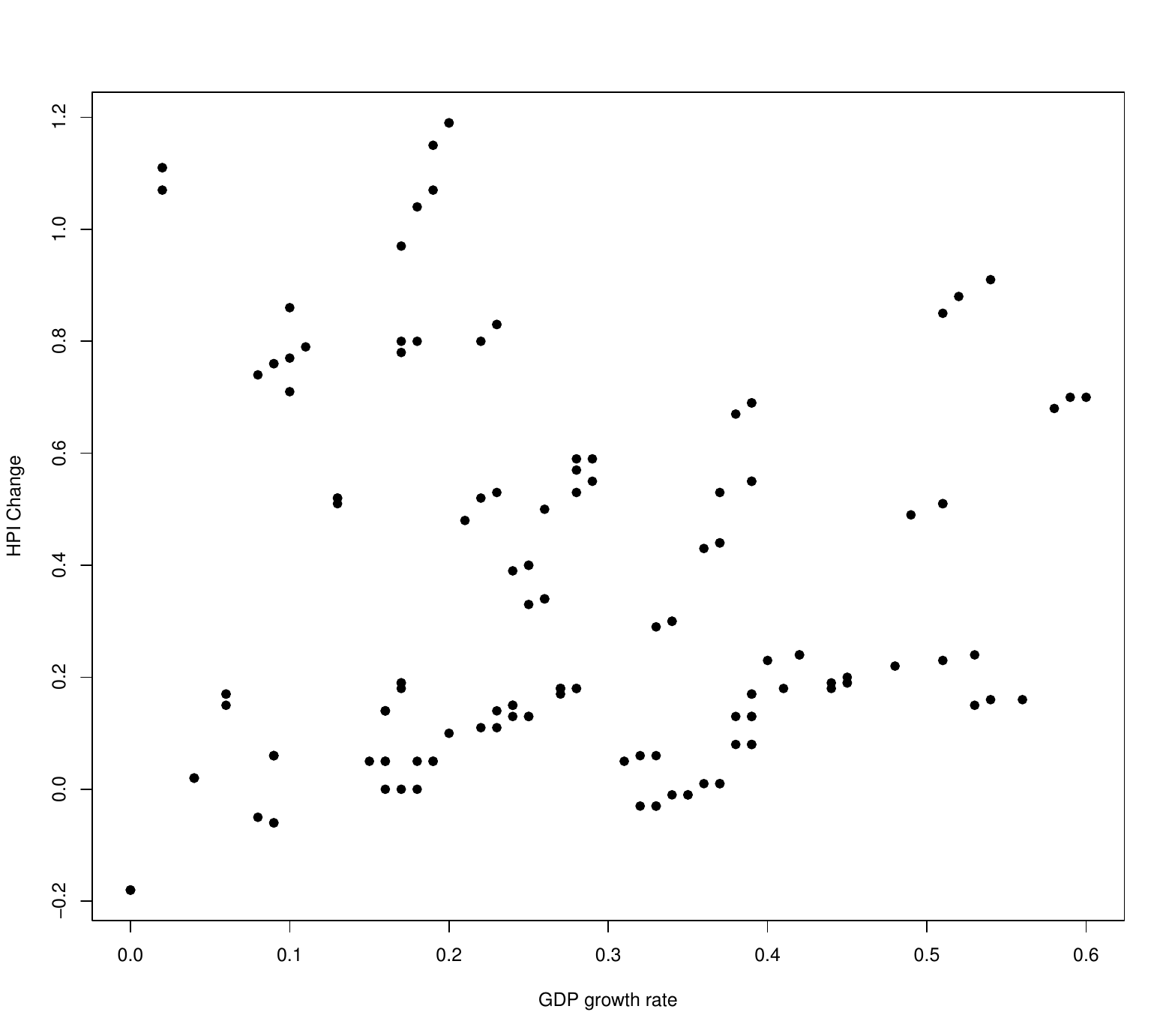}
    \caption{Scatter plot of the HPI data}
    \label{fig:hpi_data}
\end{figure}

In this section, we demonstrate the practical usefulness of the methods proposed in this paper on a real dataset that comprises of the monthly change in the house price index (HPI) and the monthly rate of growth of gross domestic product (GDP) in the United States for the period January 1990 to December 2002. The data is plotted in Figure \ref{fig:hpi_data}.

\begin{figure}[!ht]
    \centering
    \begin{subfigure}[b]{0.45\textwidth}
        \centering
        \includegraphics[width=\textwidth]{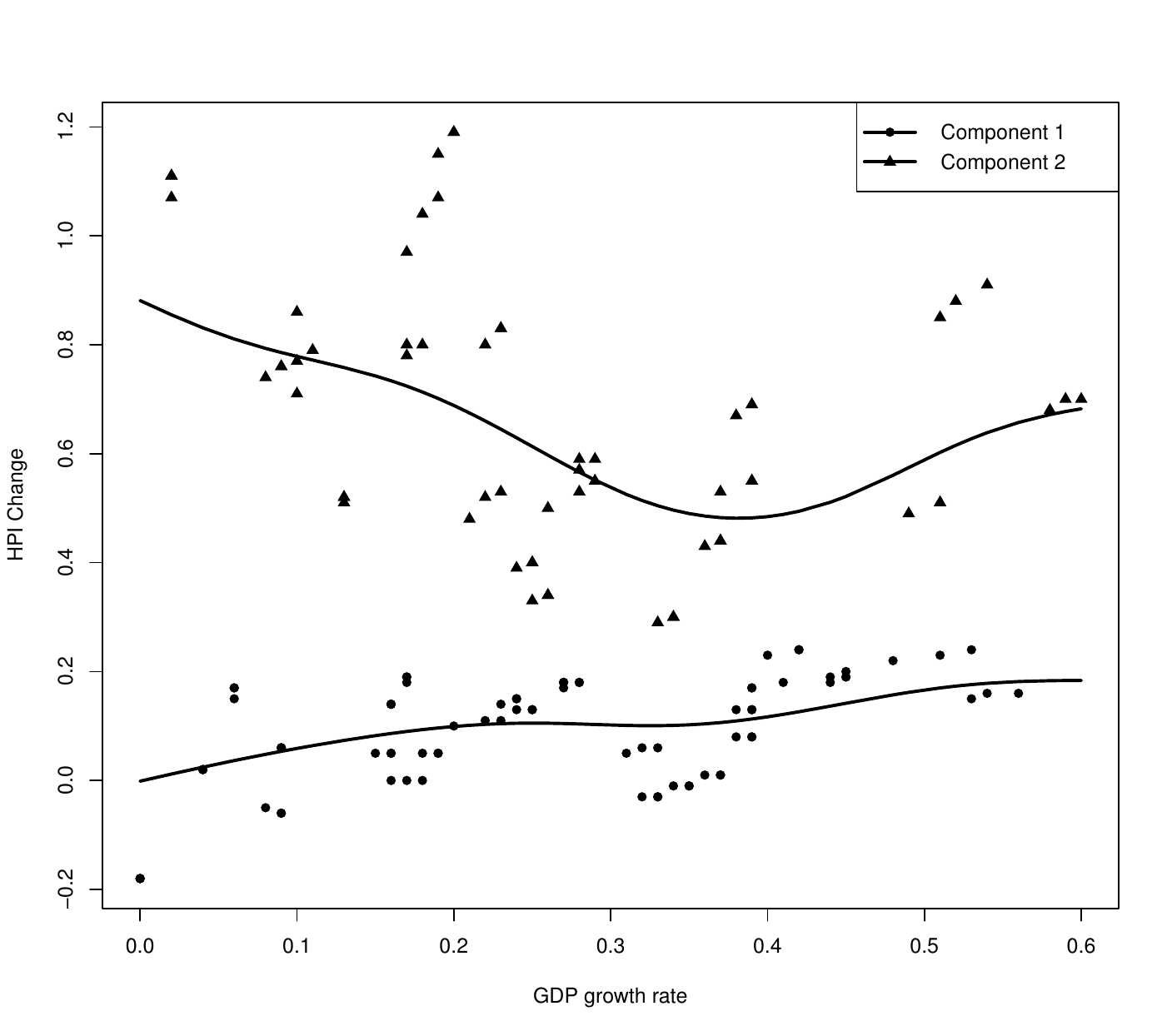}
        \caption{}
        \label{fig:sub1}
    \end{subfigure}
    \hfill
    \begin{subfigure}[b]{0.45\textwidth}
        \centering
        \includegraphics[width=\textwidth]{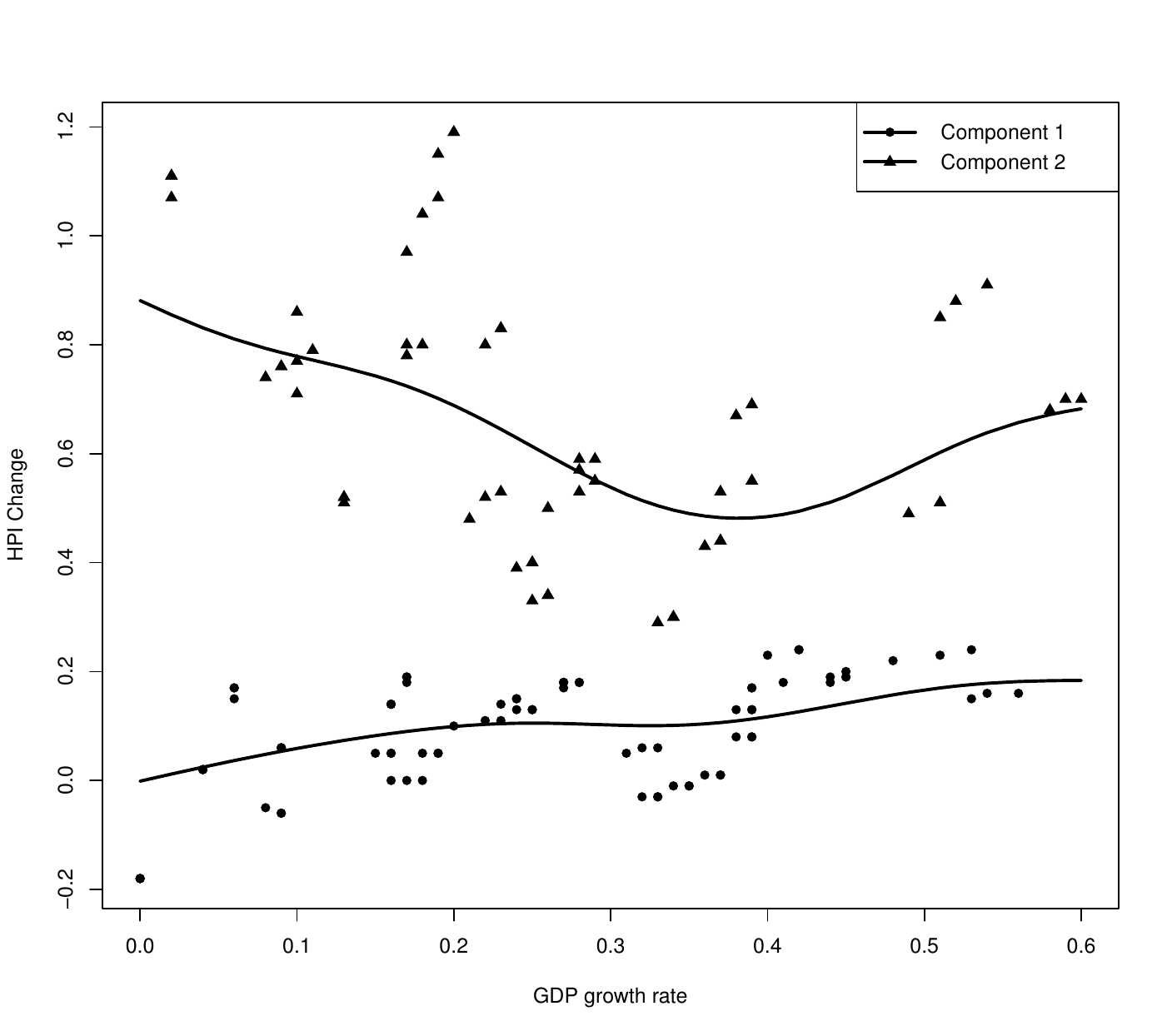}
        \caption{}
        \label{fig:sub2}
    \end{subfigure}
    \caption{The fitted component regression functions (black solid lines) for the original HPI data: (a) SPGMRs model (b) SPCGMRs model. The MAP approach was used to classify points into the two components. Points with a symbol $\bullet$ are in component 1 and points with a symbol $\blacktriangle$ are in component 2}
    \label{fig:hpi_res1}
\end{figure}

\begin{figure}[!ht]
    \centering
    \begin{subfigure}[b]{0.45\textwidth}
        \centering
        \includegraphics[width=\textwidth]{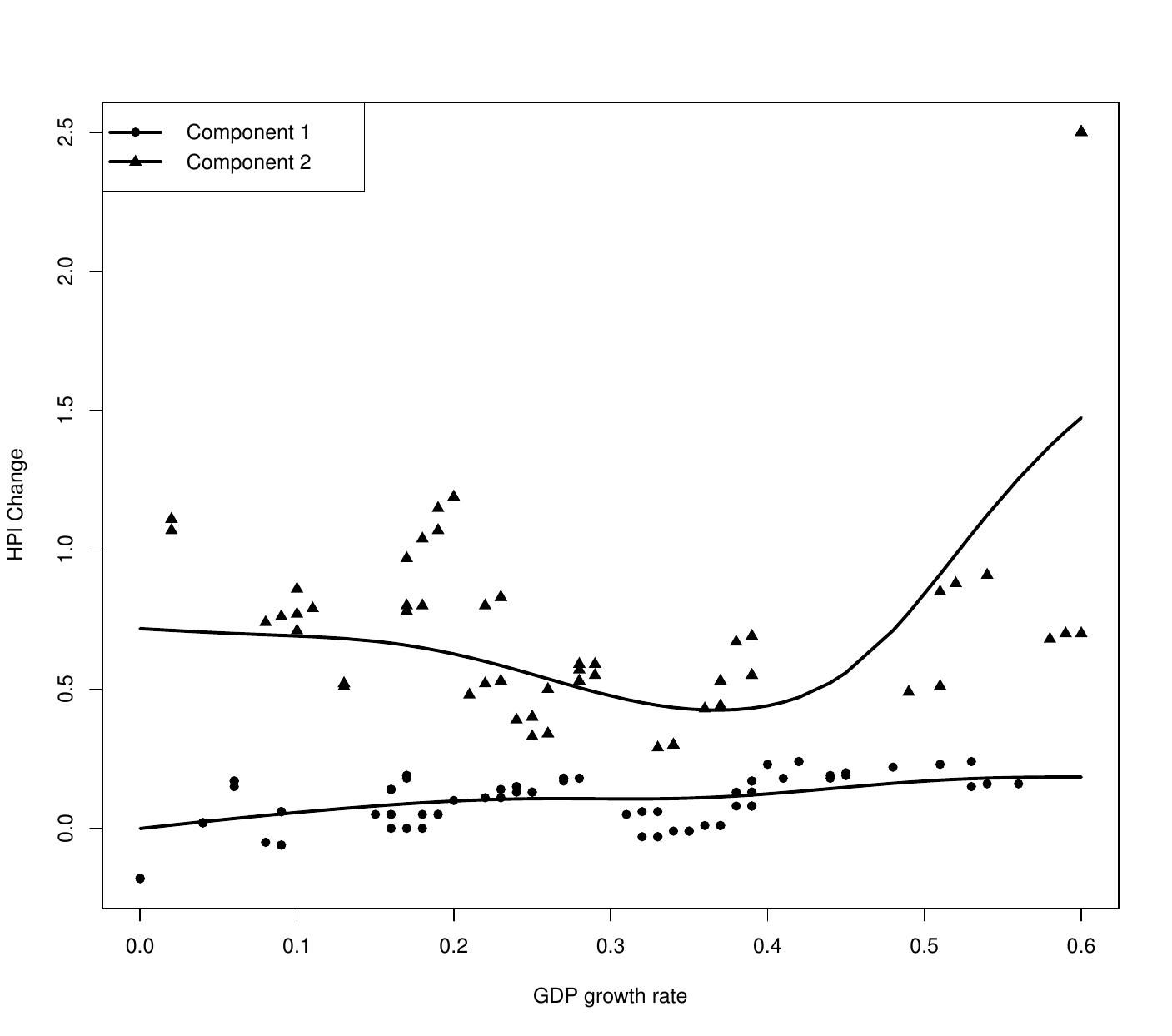}
        \caption{}
        \label{fig:sub1}
    \end{subfigure}
    \hfill
    \begin{subfigure}[b]{0.45\textwidth}
        \centering
        \includegraphics[width=\textwidth]{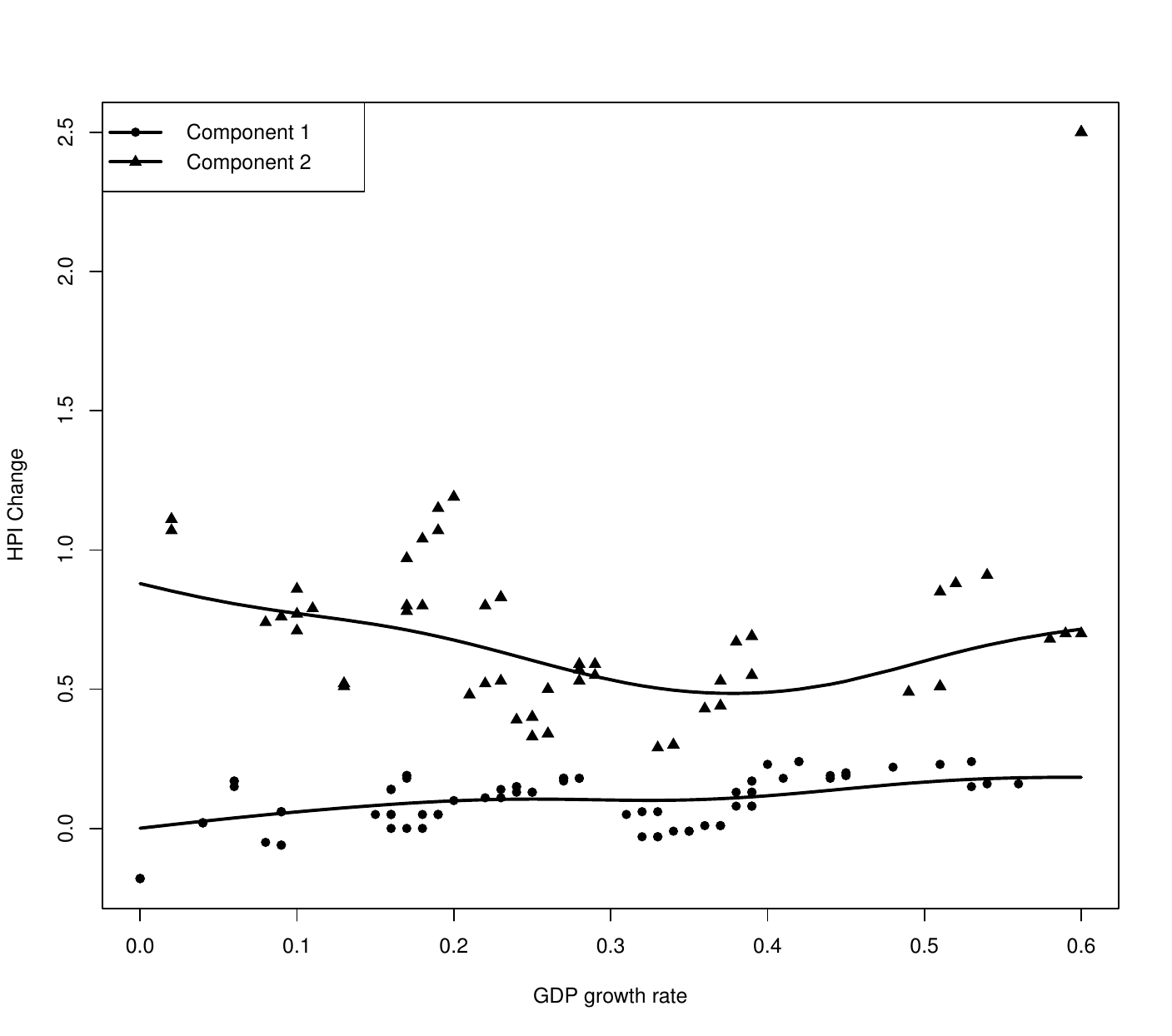}
        \caption{}
        \label{fig:sub2}
    \end{subfigure}
    \caption{The fitted component regression functions (black solid lines) for the HPI data with $5$ outliers $(0.6,2.5)$: (a) SPGMRs model (b) SPCGMRs model. The MAP approach was used to classify points into the two components. Points with a symbol $\bullet$ are in component 1 and points with a symbol $\blacktriangle$ are in component 2}
    \label{fig:hpi_out1}
\end{figure}

\begin{figure}[!ht]
    \centering
    \begin{subfigure}[b]{0.45\textwidth}
        \centering
        \includegraphics[width=\textwidth]{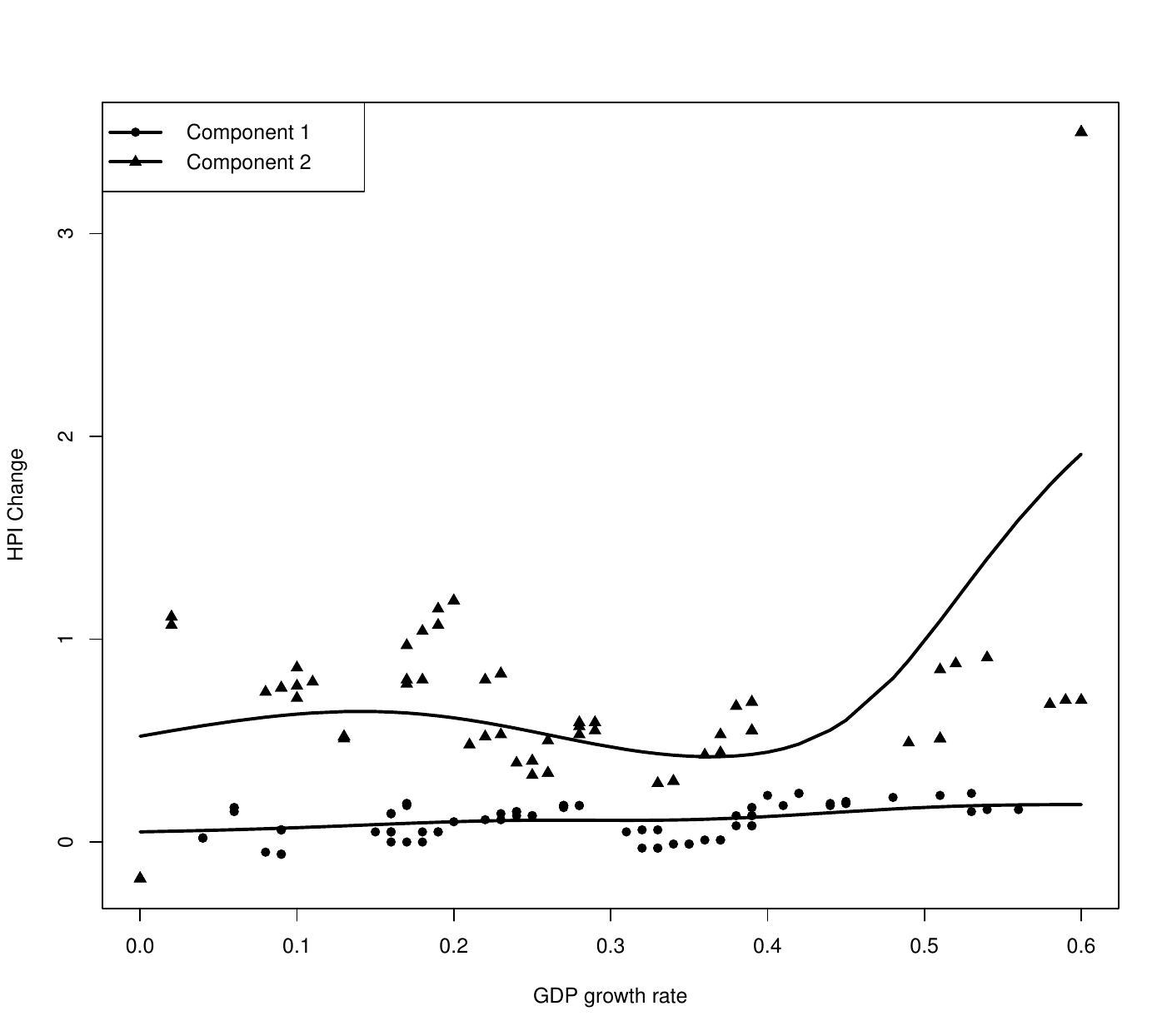}
        \caption{}
        \label{fig:sub1}
    \end{subfigure}
    \hfill
    \begin{subfigure}[b]{0.45\textwidth}
        \centering
        \includegraphics[width=\textwidth]{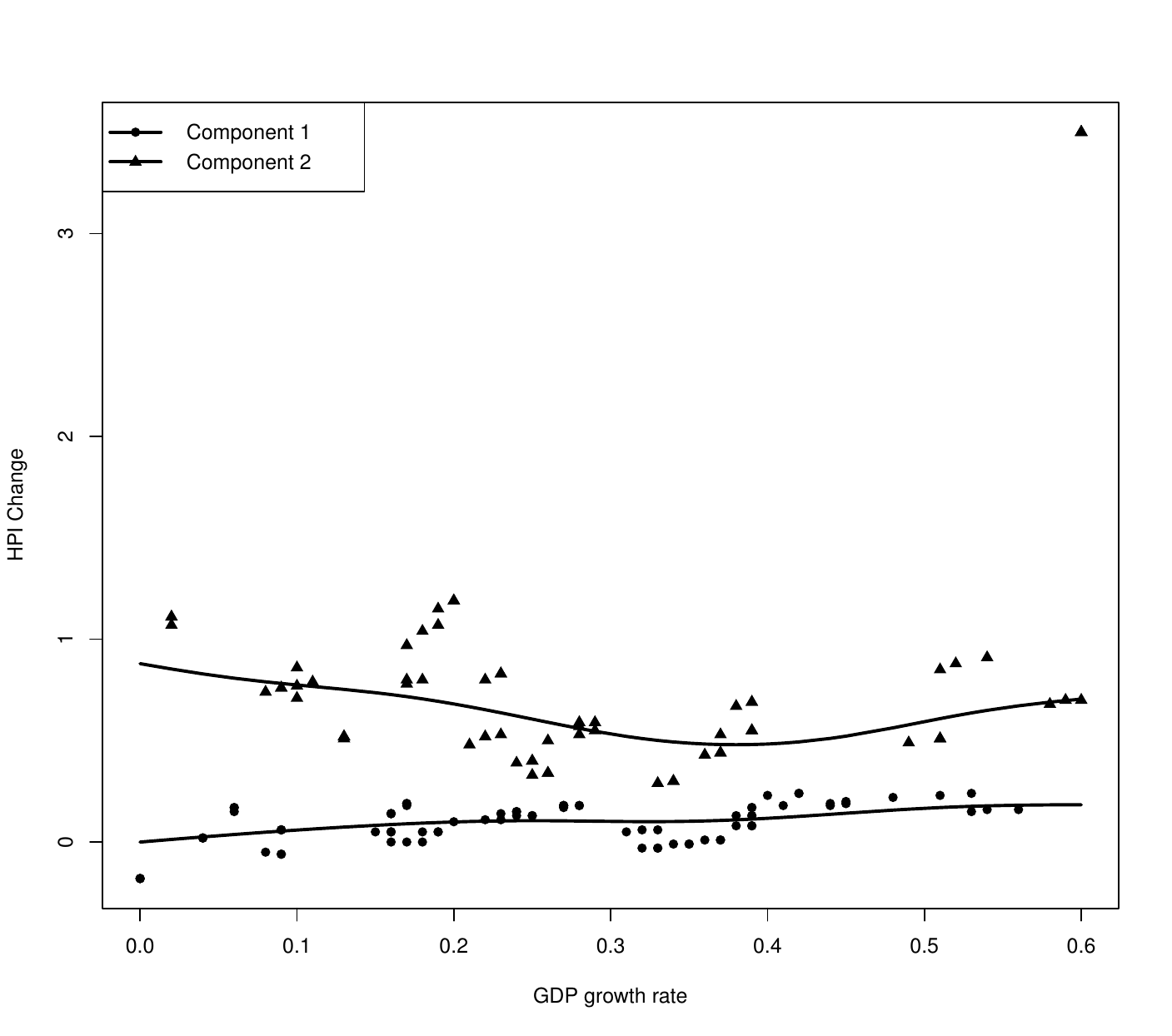}
        \caption{}
        \label{fig:sub2}
    \end{subfigure}
    \caption{The fitted component regression functions (black solid lines) for the HPI data with $5$ outliers $(0.6,3.5)$: (a) SPGMRs model (b) SPCGMRs model. The MAP approach was used to classify points into the two components. Points with a symbol $\bullet$ are in component 1 and points with a symbol $\blacktriangle$ are in component 2}
    \label{fig:hpi_out2}
\end{figure}

\begin{figure}[!ht]
    \centering
    \begin{subfigure}[b]{0.45\textwidth}
        \centering
        \includegraphics[width=\textwidth]{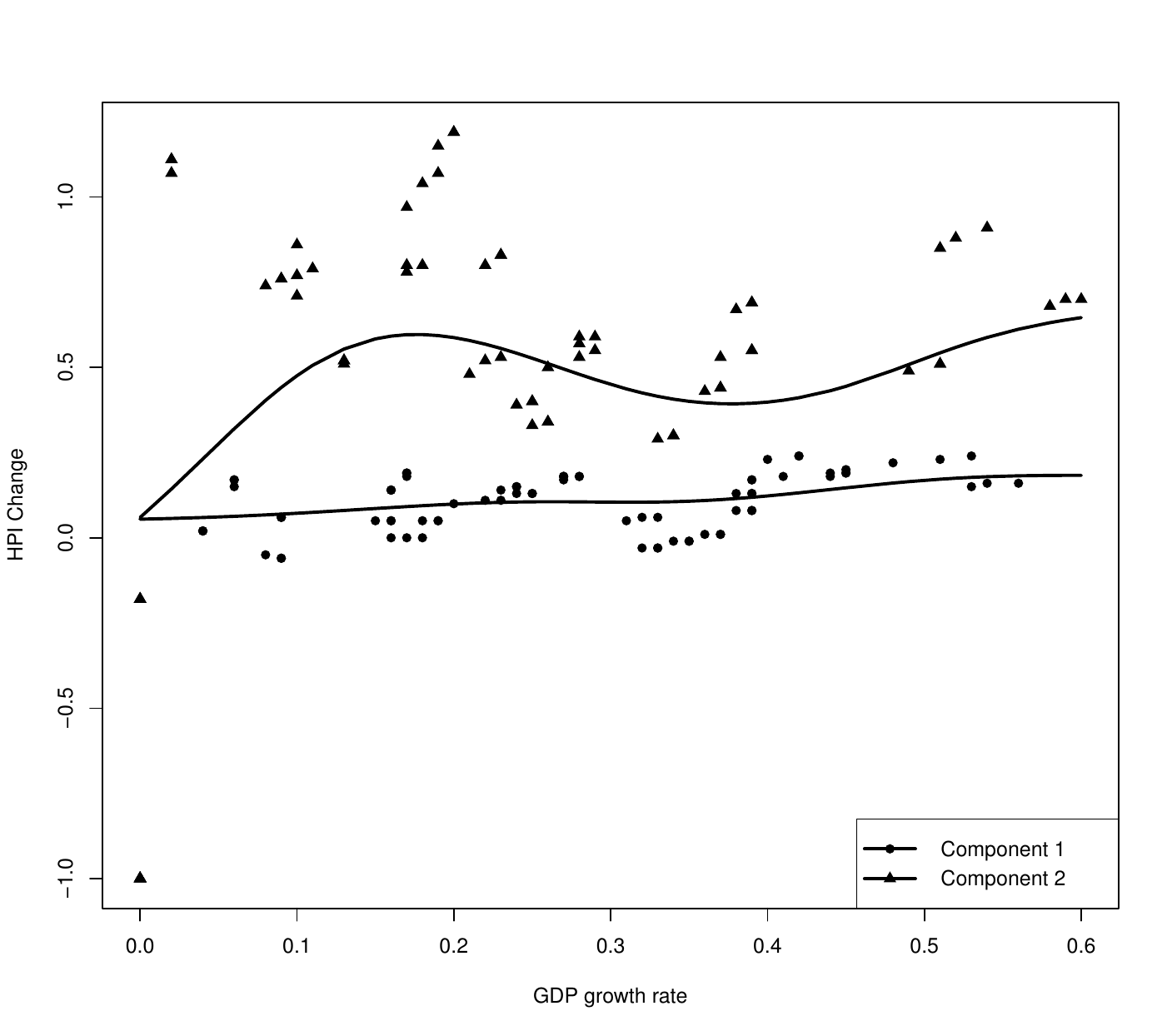}
        \caption{}
        \label{fig:sub1}
    \end{subfigure}
    \hfill
    \begin{subfigure}[b]{0.45\textwidth}
        \centering
        \includegraphics[width=\textwidth]{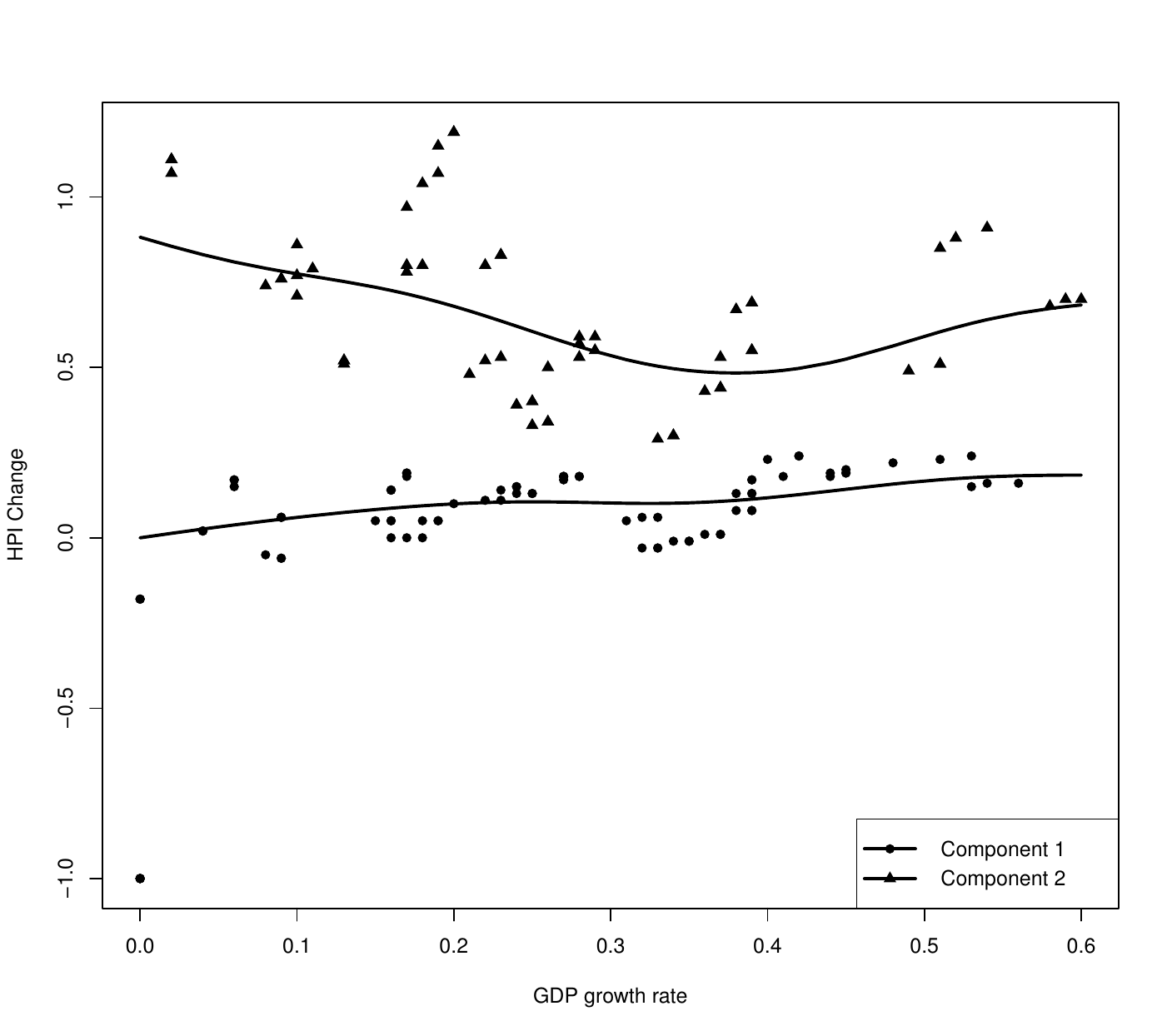}
        \caption{}
        \label{fig:sub2}
    \end{subfigure}
    \caption{The fitted component regression functions (black solid lines) for the HPI data with $5$ outliers $(0,-1)$: (a) SPGMRs model (b) SPCGMRs model. The MAP approach was used to classify points into the two components. Points with a symbol $\bullet$ are in component 1 and points with a symbol $\blacktriangle$ are in component 2}
    \label{fig:hpi_out3}
\end{figure}

It is of interest to investigate the response of HPI following a change in the GDP growth rate.  This dataset (henceforth, referred to as the HPI data) was first analysed by \cite{huang2013} using a $K=2-$component NPGMRs model. The authors interpreted the two components as two macroeconomic cycles: an economic expansion (Jan 1990 - Sep 1997) and an economic decline (Oct 1997 - Dec 2002). Later, \cite{xiang2018} fitted a $K=2$ SPGMRs model and found that their model provided the best fit to the data compared to the NPGMRs. Therefore, in this paper, we compare the performance of the SPGMRs model and the proposed SPCGMRs model on the HPI data. For $K=2$, we chose the bandwidth $h=0.75$. Table \ref{tab:hpi1} gives the estimated parameters of SPGMRs and SPCGMRs models fitted on the HPI data. Based on all the information criteria, the SPGMRs model is the best model. However, the parameter estimates from the two models are very much similar to each other. Moreover, with $\eta_1$ and $\eta_2$ close to $1$ and $\alpha_1$ and $\alpha_2$ close to $1$, we can conclude that the data is not contaminated by outliers which may be the reason why the SPGMRs model performs well. Further evidence of the similarity of the two fitted models can be seen in Figure \ref{fig:hpi_res1} which plots the fitted component regression functions obtained from the SPGMRs (left-panel) and the SPCGMRs (right-panel). The two figures are virtually the same. \\
Next, we compare the performance of the two models (SPGMRs and SPCGMRs) when the HPI data is contaminated by outliers. Following a similar approach as in \cite{song2014}, we add (a) $5$ identical pairs $(0.6,2.5)$ to the original data as outliers, (b) $5$ identical pairs $(0.6,3.5)$ and (c) $5$ identical pairs $(0,-1)$. The results of the fitted models are given in Table \ref{tab:hpi1}. It can be seen from the table that, based on all the information criteria, the SPCGMRs outperforms the SPGMRs in all the three cases. This can also be seen in Figures \ref{fig:hpi_out1} - \ref{fig:hpi_out3}, each of which plots the fitted component regression functions obtained from the SPGMRs model (left-panel) and SPCGMRs (right-panel) for all the three cases (a) - (c), respectively. It can be seen that the Gaussian SPGMRs model is grossly affected by presence of outliers whereas the SPCGMRs model appears to be unaffected. A closer look at the right-panel of Figures \ref{fig:hpi_res1} - \ref{fig:hpi_out3} reveals that the component regression functions are virtually the same. This highlights the robustness of the SPCGMRs model when the data is contaminated by mild outliers.\\
In summary, the results show that the SPGMRs model and the SPCGMRs model produce comparable results when data has no outliers. However, in the presence of mild outliers, the SPCGMRs model overwhelmingly outperforms the SPGMRs model.

\begin{sidewaystable}[!hbt]
\footnotesize
\caption{Estimated model parameters for the HPI data (bold information criteria value indicate the best model)}
\label{tab:hpi1}
\centering
\begin{tabular}{||c|c|c|c|c|c|c|c|c|c|c||}
\hline
&$\pi_1$&$\sigma^2_1$&$\sigma^2_2$&$\alpha_1$&$\alpha_2$&$\eta_1$&$\eta_2$&AIC&BIC&ICL\\
\hline 
&\multicolumn{10}{|c||}{Original data}\\
\hline
SPGMRs&0.5469&0.0066&0.0460&-&-&-&-&\textbf{15.7154}&\textbf{69.4718}&\textbf{80.0856}\\
SPCGMRs&0.5469&0.0066&0.0460&0.9983&0.9967&1.0001&1.0001&23.7154&89.2667&99.8806\\
\hline
&\multicolumn{10}{|c||}{Original data with $5$ outliers $(0.6,2.5)$}\\
\hline
SPGMRs&0.4709&0.0061&0.1660&-&-&-&-&95.2948&149.6864&178.5223\\
SPCGMRs&0.5257&0.0066&0.0428&0.9304&0.8822&1.0001&47.4332&\textbf{83.3529}&\textbf{149.6789}&\textbf{160.854}\\
\hline
&\multicolumn{10}{|c||}{Original data with $5$ outliers $(0.6,3.5)$}\\
\hline
SPGMRs&0.4464&0.0051&0.3108&-&-&-&-&143.288&197.6797&228.1139\\
SPCGMRs&0.5269&0.0066&0.0445&0.9977&0.9061&1.0001&136.8506&\textbf{89.4892}&\textbf{155.8152}&\textbf{166.7284}\\
\hline
&\multicolumn{10}{|c||}{Original data with $5$ outliers $(0,-1)$}\\
\hline
SPGMRs&0.4098&0.0047&0.1670&-&-&-&-&92.8734&147.265&189.3214\\
SPCGMRs&0.5887&0.0064&0.0433&0.8771&0.9987&100.1901&1.0001&\textbf{74.8567}&\textbf{141.1827}&\textbf{158.063}\\
\hline
\end{tabular}
\end{sidewaystable}

\section{Conclusion}\label{conc}
In this paper, we proposed robust and flexible mixtures of regressions referred to as non-parametric contaminated Gaussian mixtures of regressions (NPCGMRs). The NPCGMRs model assumes that (1) the mixing proportions, component regressions functions and variances are smooth unknown, and hence non-parametric, functions of the covariates and (2) each component error distribution follows a contaminated Gaussian distribution. Compared to the traditional non-parametric Gaussian mixtures of regressions (NPGMRs) \cite{huang2013}, which assume that each component error distribution follows a Gaussian distribution, the NPCGMRs is less sensitive to the presence of mild outliers in the data. This was demonstrated through an extensive simulation study and a sensitivity analysis on a real data set.\\
It is important to mention that the scenarios considered for our numerical studies, were only concerned with mild outliers. That is, data points that renders the component conditional distribution of the response variable $y$ to be heavy-tailed. That is to say, we only considered outliers in the $y$ values. Future studies should study the performance of the proposed model in scenarios in which outliers may be present in both the response variable and the covariate. Note that, to avoid the well-known curse-of-dimensionality problem in non-parametric kernel estimation, the proposed model was introduced for the case of one covariate. However, the model can be easily extended to accommodate more than one covariate by using additive models \cite{zhang2018}, single-index models \cite{xiang2020} or neural networks \cite{xue2025}.

\section*{Declarations}
Conflict of interest/Competing interests: The authors declare no conflict of interest.
\bibliography{References}

@article{biernacki2000,
  title={Assessing a mixture model for clustering with the integrated completed likelihood},
  author={Biernacki, Christophe and Celeux, Gilles and Govaert, G{\'e}rard},
  journal={IEEE transactions on pattern analysis and machine intelligence},
  volume={22},
  number={7},
  pages={719--725},
  year={2000}
}

@article{xue2025,
  title={{Machine learning embedded EM algorithms for semiparametric mixture regression models}},
  author={Xue, J. and Yao, W. and Xiang, S.},
  journal={Computational Statistics},
  volume={40},
  number={1},
  pages={205--224},
  year={2025}
}

@article{meng1993,
  title={{Maximum likelihood estimation via the ECM algorithm: A general framework}},
  author={Meng, X. and Rubin, D. B.},
  journal={Biometrika},
  volume={80},
  number={2},
  pages={267--278},
  year={1993}
}

@article{quandt1972,
  title={A new approach to estimating switching regressions},
  author={Quandt, R. E.},
  journal={Journal of the American Statistical Association},
  volume={67},
  number={338},
  pages={306--310},
  year={1972}
}

@article{quandt1978,
  title={Estimating mixtures of normal distributions and switching regressions},
  author={Quandt, R. E. and Ramsey, J. B.},
  journal={Journal of the American Statistical Association},
  volume={73},
  number={364},
  pages={730--738},
  year={1978}
}

@article{xiang2018,
  title={Semiparametric mixtures of nonparametric regressions},
  author={Xiang, S. and Yao, W.},
  journal={Annals of the Institute of Statistical Mathematics},
  year={2018},
  doi={10.1007/s10463-016-0584-7}
}

@article{zhang2018,
  title={Semiparametric mixture of additive regression models},
  author={Zhang, Y. and Zheng, Q.},
  journal={Communications in Statistics-Theory and Methods},
  year={2018},
  doi={10.1080/03610926.2017.1310243}
}

@article{yu2020,
  title={A selective overview and comparison of robust mixture regression estimators},
  author={Yu, C. and Yao, W. and Yang, G.},
  journal={International Statistical Review},
  volume={88},
  number={1},
  pages={176--202},
  year={2020}
}

@article{xiang2020,
  title={Semiparametric mixtures of regressions with single-index for model based clustering},
  author={Xiang, S. and Yao, W.},
  journal={Advances in Data Analysis and Classification},
  year={2020},
  doi={10.1007/s11634-020-00392-w}
}

@inproceedings{shen2004,
  title={Outlier detecting in fuzzy switching regression models},
  author={Shen, H. and Yang, J. and Wang, S.},
  booktitle={Artificial Intelligence: Methodology, Systems, and Applications: 11th International Conference, AIMSA 2004, Varna, Bulgaria, September 2-4, 2004. Proceedings 11},
  pages={208--215},
  year={2004}
}

@article{garcia2017,
  title={Robust estimation of mixtures of regressions with random covariates, via trimming and constraints},
  author={Garc{\'\i}a-Escudero, L. A. and Gordaliza, A. and Greselin, F. and Ingrassia, S. and Mayo-{\'I}scar, A.},
  journal={Statistics and Computing},
  volume={27},
  pages={377--402},
  year={2017}
}

@article{wang2014,
  title={A note on the identifiability of nonparametric and semiparametric mixtures of GLMs},
  author={Wang, S. and Yao, W. and Huang, M.},
  journal={Statistics \& Probability Letters},
  volume={93},
  pages={41--45},
  year={2014}
}

@article{punzo2016,
  title={Parsimonious mixtures of multivariate contaminated normal distributions},
  author={Punzo, A. and McNicholas, P. D.},
  journal={Biometrical Journal},
  volume={58},
  number={6},
  pages={1506--1537},
  year={2016}
}

@article{zeller2016,
  title={Robust mixture regression modeling based on scale mixtures of skew-normal distributions},
  author={Zeller, C. B. and Cabral, C. R. B. and Lachos, V. H.},
  journal={Test},
  volume={25},
  pages={375--396},
  year={2016}
}

@article{bai2012,
  title={Robust fitting of mixture regression models},
  author={Bai, X. and Yao, W. and Boyer, J. E.},
  journal={Computational Statistics \& Data Analysis},
  volume={56},
  number={7},
  pages={2347--2359},
  year={2012}
}

@article{markatou2000,
  title={Mixture models, robustness, and the weighted likelihood methodology},
  author={Markatou, M.},
  journal={Biometrics},
  volume={56},
  number={2},
  pages={483--486},
  year={2000}
}

@article{song2014,
  title={{Robust mixture regression model fitting by Laplace distribution}},
  author={Song, Weixing and Yao, Weixin and Xing, Yanru},
  journal={Computational Statistics \& Data Analysis},
  volume={71},
  pages={128--137},
  year={2014}
}

@article{dougru2018,
  title={Robust mixture regression modeling using the least trimmed squares (lts)-estimation method},
  author={Do{\u{g}}ru, F. Z. and Arslan, O.},
  journal={Communications in Statistics-Simulation and Computation},
  volume={47},
  number={7},
  pages={2184--2196},
  year={2018}
}

@article{neykov2007,
  title={Robust fitting of mixtures using the trimmed likelihood estimator},
  author={Neykov, N. and Filzmoser, P. and Dimova, R. and Neytchev, P.},
  journal={Computational Statistics \& Data Analysis},
  volume={52},
  number={1},
  pages={299--308},
  year={2007}
}

@article{ge2024,
  title={t-distribution-based robust semiparametric mixture regression model},
  author={Ge, Y. and Xiang, S. and Yao, W.},
  journal={Communications in Statistics-Simulation and Computation},
  pages={1--21},
  year={2024}
}

@article{yao2014,
  title={Robust mixture regression using the t-distribution},
  author={Yao, W. and Wei, Y. and Yu, C.},
  journal={Computational Statistics \& Data Analysis},
  volume={71},
  pages={116--127},
  year={2014}
}

@article{skhosana2024,
  title={{A modified EM-type algorithm to estimate semi-parametric mixtures of non-parametric regressions}},
  author={Skhosana, S. B. and Millard, S. M. and Kanfer, F. H. J.},
  journal={Statistics and Computing},
  volume={34},
  number={4},
  pages={125},
  year={2024}
}

@article{wu2017,
  title={Estimation and testing for semiparametric mixtures of partially linear models},
  author={Wu, X. and Liu, T.},
  journal={Communications in Statistics-Theory and Methods},
  year={2017},
  doi={10.1080/03610926.2016.1189569}
}

@article{huang2013,
  title={Nonparametric mixture of regression models},
  author={Huang, M. and Li, R. and Wang, S.},
  journal={Journal of the American Statistical Association},
  year={2013},
  doi={10.1080/01621459.2013.772897}
}

@article{huang2012,
  title={Mixture of regression models with varying mixing proportions: a semiparametric approach},
  author={Huang, M. and Yao, W.},
  journal={Journal of the American Statistical Association},
  year={2012},
  doi={10.1080/01621459.2012.682541}
}

@article{tibshirani1987,
  title={Local likelihood estimation},
  author={Tibshirani, R. and Hastie, T.},
  journal={Journal of the American Statistical Association},
  volume={82},
  number={398},
  pages={559--567},
  year={1987}
}

@article{yao2012,
  title={Model based labeling for mixture models},
  author={Yao, W.},
  journal={Statistics and Computing},
  volume={22},
  pages={337--347},
  year={2012}
}

@article{schwarz1978,
 author = {Gideon Schwarz},
 journal = {The Annals of Statistics},
 number = {2},
 pages = {461--464},
 publisher = {Institute of Mathematical Statistics},
 title = {Estimating the Dimension of a Model},
 volume = {6},
 year = {1978}
}

@article{akaike1974,
  author={Akaike, H.},
  journal={IEEE Transactions on Automatic Control}, 
  title={A new look at the statistical model identification}, 
  year={1974},
  volume={19},
  number={6},
  pages={716-723},
  doi={10.1109/TAC.1974.1100705}}

@book{frühwirth2006,
  title={{Finite Mixture and Markov Switching Models}},
  author={Fr{\"u}hwirth-Schnatter, S.},
  series={Springer Series in Statistics},
  year={2006},
  address={New York, NY},
  publisher={Springer}
}

@article{desarbo1988,
  title={A maximum likelihood methodology for clusterwise linear regression},
  author={DeSarbo, W. S. and Cron, W. L.},
  journal={Journal of Classification},
  volume={5},
  pages={249--282},
  year={1988}
}

@book{yao2024,
  title={Mixture Models: Parametric, Semiparametric, and New Directions},
  author={Yao, W. and Xiang, S.},
  year={2024},
  address={Boca Raton},
  publisher={CRC Press}
}

@article{jacobs1991,
  title={Adaptive mixtures of local experts},
  author={Jacobs, R. A. and Jordan, M. I. and Nowlan, S. J. and Hinton, G. E.},
  journal={Neural Computation},
  volume={3},
  number={1},
  pages={79--87},
  year={1991}
}

@book{schlattmann2009,
  title={Medical applications of finite mixture models.},
  author={Schlattmann, P.},
  year={2009},
  address={Berlin, Heidelberg},
  publisher={Springer}
}

@article{mazza2020,
  title={Mixtures of multivariate contaminated normal regression models},
  author={Mazza, A. and Punzo, A.},
  journal={Statistical Papers},
  volume={61},
  number={2},
  pages={787--822},
  year={2020}
}

@article{stephens2000,
  title={Dealing with label switching in mixture models},
  author={Stephens, M.},
  journal={Journal of the Royal Statistical Society: Series B (Statistical Methodology)},
  volume={62},
  number={4},
  pages={795--809},
  year={2000}
}

@article{DLR1977,
  title={{Maximum likelihood from incomplete data via the EM algorithm}},
  author={Dempster, A. P. and Laird, N. M. and Rubin, D. B.},
  journal={Journal of the Royal Statistical Society: Series B (Methodological)},
  volume={39},
  number={1},
  pages={1--22},
  year={1977}
}
\bibliographystyle{plainnat}

\end{document}